\documentclass[aps,pre,
				superscriptaddress,
				twocolumn,balancelastpage,nofootinbib]{revtex4-2}
\usepackage{amsmath}
\usepackage{amssymb}
\usepackage{graphicx}
\usepackage{bbold}
\usepackage{color}
\usepackage[colorlinks=true,
urlcolor=blue,
linkcolor=blue,
citecolor=blue]{hyperref}
\usepackage[all]{hypcap}
\usepackage{url}
\newcommand{\kt}[1]{\left \lvert #1 \right \rangle}
\newcommand{\br}[1]{\left \langle #1 \right \rvert}
\newcommand{\rev}[1]{{\textcolor{black}{ #1}}} 
\DeclareMathOperator{\tr}{tr}

\graphicspath{{../figures/}}
\newcommand{\unity}{\mathbb{1}}
\begin{document}
\title{Sunburst {\color{black} quantum Ising} model under interaction 
quench: entanglement and role of initial state coherence}
\author{Akash Mitra}
\author{Shashi C. L. Srivastava}
\affiliation{Variable Energy Cyclotron Centre, 1/AF Bidhannagar, Kolkata 
	700064, India}
\affiliation{Homi Bhabha National Institute, Training School Complex, 
	Anushaktinagar, Mumbai - 400094, India}
\begin{abstract}
	We study the non-equilibrium dynamics of an isolated bipartite 
	quantum 
	system, the sunburst quantum Ising model, under interaction 
	quench. The pre-quench limit of this model is two non-interacting	
	integrable systems, namely a transverse ising chain and 
	finite number of isolated qubits. As a function of interaction 
	strength,  the spectral fluctuation property goes from Poisson to 
	Wigner-Dyson statistics. We chose entanglement entropy as a probe 
	to 
	study the approach to thermalization or lack of it in post-quench 
	dynamics. In the near-integrable limit, as expected, the linear 
	entropy displays oscillatory behavior while in the chaotic limit, 
	it 
	saturates. Along with the chaotic nature of the time evolution 
	generator, we show the importance of the role played by the 
	coherence 
	of the initial state in deciding the nature of thermalization. We 
	further show that these findings are general by replacing the 
	Ising 
	ring with a disordered $XXZ$ model with disorder strength putting 
	it in 
	the many-body localized phase.
\end{abstract}
\maketitle

\section{Introduction}
There has been an upsurge of studies in the last decade, both 
theoretically and experimentally, to understand fundamental questions 
like relaxation dynamics and thermalization in out-of-equilibrium 
isolated quantum systems. Traditional quantifiers like two-point or higher 
{\color{black}order} correlations of local operators and quantum 
information theoretic measures like entanglement entropy and out-of-time 
ordered correlators, among others, have been used as probes to understand 
whether the system attains the steady state, what is the nature of 
equilibrium or what is the path to achieve the same. The equilibrium state 
is described by Generalized Gibbs ensemble (GGE) or eigenstate 
thermalization hypothesis (ETH) depending on the system being integrable 
or quantum 
chaotic\cite{Kitowe_2006,Cheneau_2012,Langen_2015,Brydges_2019,Lukin_2019,Schneider_2012,Gring_2012,Kaufman_2016,Trotzky_2012,Langen_2013,Islam2015,Ritter_2007}.
ETH essentially asserts that each energy eigenstate behaves like a 
micro-canonical ensemble \cite{Deutsch_1991,Srednicki_1994,Deutsch_2018}. 
Several analytical and numerical studies have shown that generic isolated 
quantum chaotic many-body systems relax to a state in which local 
observables achieve a time-independent value predicted by thermal 
distribution and are independent of the initial state as long as they are 
chosen from a small band of considered 
energy\cite{Rigol_2008,Rigol_2012,Rigol_2016,Santos_2010,Santos_2013}.
A generic approach for creating a system that does not thermalize is 
provided by many-body localization (for further information, see the 
reviews \cite{Rahul_2015, AleLaf2018, Abanin_2019} and references therein).
Another notable exception occurs when a many-body interacting system 
exhibits persistent revivals for specific initial states from the 
middle of the spectrum while exhibiting ergodicity for others. These 
specific initial states are called many-body scarred states 
\cite{BerSchKeeLev2017,  BluOmrKeeSemetal2021}. 	

The dynamics of bipartite subsystem entanglement entropy following a 
quantum quench that drives the system away from its initial equilibrium 
state provides an equivalent method for studying thermalization in 
many-body systems. The Von Neumann entropy is one of the most 
commonly used entanglement measures in quantum information theory; 
however, we choose linear entropy for the ease of the calculation as 
well as being accessible to experimental measurements 
\cite{Islam2015}. A many-body system can be imagined as a bipartite system 
with interaction strength between the two parts as a natural 
candidate for the quench parameter. {\color{black}The quench protocol 
	for Hamiltonian is written as,
	\begin{equation}\label{eq:bipartite}
		H = \begin{cases}
			H_A \otimes \unity_{B} + \unity_{A} \otimes H_B & t<0\\
			H_A \otimes \unity_{B} + \unity_{A} \otimes H_B + V_{AB}& 
			t\geq 0
		\end{cases}
	\end{equation}
	where $H_{A(B)}$ is Hamiltonian of subsystem $A(B)$ and $V_{AB}$ is 
	the interaction between the two. Time is denoted by $t$.
	
	When $H_{A}, H_B$, as well as post-quench Hamiltonian $H$, is 
	integrable, the analytical understanding of entanglement evolution 
	comes from  quasiparticle picture \cite{Calabrese_2005}. A 
	phenomenological model which rather successfully predicts the 
	entanglement between two subsystems at time $t$ in terms of the 
	number of intersecting trajectories of quasiparticles produced at 
	time $t=0$ in two subsystems. The entanglement entropy grows linearly 
	in time before saturating to volume law entanglement
	\cite{Calabrese_2016,Eisler_2008,Alba_2018,Chiara_2006,Calabrese_2020,Fagotti_2008,Eisler_2008,Nezgharaja_2014,Coser_2014,Buyantmauessfaband_2016,Kudler_Flam_2021,Alba_2017,Alba_2021}.
	In the other scenario, when $H_A, H_B$, as well as post-quench 
	Hamiltonian $H$, are quantum chaotic, a random matrix theory-based 
	approach explains the entanglement dynamics \cite{Arul_2016}. The 
	quantum chaotic nature of subsystems is crucial in this approach as 
	all the fluctuation and entanglement properties have been calculated 
	by modeling subsystem Hamiltonian with suitable random matrix 
	ensemble 
	\cite{Arul_2001,Arul_2002,Fujisaki_2003,Shashi_2016,Arul_2016,Steven_2018,
	 Pulikkottil_2020}. It motivates us to study the third scenario where 
	pre-quench subsystems are integrable and interaction quench breaks the 
	integrability of the post-quench system. The present study departs 
	from the two previously studied scenarios as the integrability of 
	subsystems forbids a random matrix theory-based approach. However, we 
	can also not use a quasiparticle picture due to the quantum chaotic 
	nature of the post-quench Hamiltonian. For this class of systems in 
	which the total system is quantum chaotic due to coupling between the 
	two 
	integrable systems, it is a valid question to ask whether the 
	equilibrium state of any of the two subsystems behaves like a thermal 
	state or it can be characterized by GGE as is expected for integrable 
	systems.}

For the second scenario discussed above, where subsystems are modeled by 
corresponding random matrix ensemble, the initial coherence of the quantum 
state\rev{, defined as the sum of the square of off-diagonal elements of 
the density matrix written in energy basis,} under interaction quench is 
shown to act as a resource for 
equilibration or thermalization even when the ETH is not 
satisfied \cite{Arul_2022}. The equilibration is called strong or 
weak depending on whether the initial state averaged temporal 
fluctuation of linear entropy is small or not. The equilibration will be 
referred to as  thermalization whenever the equilibrium value of linear 
entropy reaches the corresponding random vector value, henceforth called 
Lubkin value\cite{Lub1978}. {\color{black} For a random 
	vector chosen uniformly from Hilbert space $\mathcal{H}_A \otimes 
	\mathcal{H}_B$ of dimension $2^{L+n}$  with dimension of 
	$\mathcal{H}_{A(B)}$, $2^{L}(2^n)$ such that $L>>n$, the linear 
	entropy 
	of the either subsystem is $\approx 1- \frac{1}{2^n}$}.

In this paper, we study the spectral fluctuation of the static  
sunburst quantum Ising model \cite{Franchi_2022} and show a transition 
from 
Poisson 
to Wigner-Dyson statistics as a function of interaction strength. Then, 
using linear entropy as a probe, we study the entanglement dynamics of 
qubits under interaction quench. We also show how initial state coherence 
plays a role in achieving strong thermalization under interaction quench. 
The key differences from the existing literature are as follows:
\begin{itemize}
	\item In the bipartite setting shown in Eq. \ref{eq:bipartite}, 
	the pre-quench limit of the \textit{quantum chaotic} system is two 
	non-interacting integrable subsystems in contrast to the 
	non-integrable limit studied in the literature \cite{Arul_2001, 
		Arul_2002, Fujisaki_2003, Arul_2016, Arul_2022}.
	\item The coherence of the initial state is shown to 
	control the mean and variance of the post-quench equilibrium 
	value of the linear entropy.
\end{itemize}
The paper is organized as follows: We recollect the details of the 
sunburst quantum Ising model and results relevant to the present work in 
Section \ref{sec:model}, then study the transition from integrability to 
chaos with varying interaction strength between the Ising ring and qubits 
in Section \ref{sec:transition}. We discuss the post-quench entanglement 
dynamics following the interaction quench in Section \ref{sec:quench}, 
while Section \ref{sec:coherence} focuses on the role of the initial state 
coherence in controlling the nature of thermalization. We present the 
summary and outlook in Section \ref{sec:summary}.

\section{Model}\label{sec:model}
We study {\color{black}a} recently proposed sunburst {\color{black}quantum 
Ising} model that is composed of a transverse field Ising ring, 
symmetrically coupled to a few isolated external qubits 
\cite{Franchi_2022}. {\color{black}The subsystems $A$ and $B$ in 
Eq.\ref{eq:bipartite} correspond to the transverse field Ising model and 
isolated qubits, respectively.} The total Hamiltonian of this model is 
\begin{equation}\label{eq:model_H}
	H = H_I \otimes \unity_{q} + \unity_{I} \otimes H_q + V_{Iq},
\end{equation}
where $ H_I, H_q $ are Hamiltonian of transverse field Ising ring, 
isolated qubits respectively and $ V_{Iq} $ is the interaction term. 
$ \unity_{q(I)} $ is the identity operator in the space of qubits 
(Ising ring). The individual terms are defined as,
\begin{equation}\label{eq:subsystems}
	\begin{aligned}
		H_I&=-\sum_{i=1}^L \left(J\sigma_i^x \sigma_{i+1}^x 
		+h\sigma_i^z\right)\\
		H_q &= -\frac{\delta}{2}\sum_{i=1}^n \Sigma_i^z; 
		\quad 
		{\color{black}	V_{Iq} = -\kappa \sum_{i=1}^n \sigma_{1+(i-1)b}^x 
			\Sigma_{i}^x}
	\end{aligned}
\end{equation}
where $L$ is the number of lattice points in the Ising ring and 
$\sigma_i^{x(z)}$ denotes Pauli matrices on the $i$th site. Note that 
both subsystems are integrable. Due to the ring topology of Ising, 
$\sigma_{L+1}^x=\sigma_1^x$.  $ J $ is hopping strength which we 
choose as unity unless stated otherwise, and $h$ is the strength of the 
transverse field.
The spin chain is coupled with $n$ number of isolated qubits for 
which the Hamiltonian is $ H_q $ and $\Sigma_i$ denotes the Pauli 
matrix corresponding to $i$th qubit. The energy gap between the two 
lowest eigenstates for the isolated qubits is denoted by $\delta$. $ 
\kappa $ is the strength of homogeneous interaction between a qubit 
and Ising spin {\color{black}while}  $b$ represents the distance 
between 
consecutive isolated qubits. For the case $ L = nb $, the complete 
Hamiltonian is translation invariant with a unit cell containing $ b 
$ Ising spins and one isolated qubit. For all the other cases, this 
symmetry is broken. 

In addition to translation symmetry, the model has spin-flip symmetry, 
i.e. the Hamiltonian remains invariant when spin is flipped along $x$ 
and $y$ direction while keeping spin in $z$ direction unchanged. The 
symmetry operator is given by,
\begin{equation}\label{eq:parity}
	P=\displaystyle{\prod_{i=1}^L 
		\sigma_i^z}\otimes\displaystyle{\prod_{j=1}^n \Sigma_j^z}
\end{equation}
which commutes with the Hamiltonian in Eq. \ref{eq:model_H} and 
satisfies $ P^2 = I $. Therefore, Hamiltonian has a disjoint spectrum 
corresponding to $ P = \pm 1 $. For integrable to chaotic transition, 
spectral distribution is studied for a fixed symmetry sector, 
{\color{black}$P=+1$}. 

{\color{black}To show the generality of results obtained in subsequent 
	sections, we also study a version of the sunburst 
	{\color{black}quantum 
		Ising }model where we 
	replace transverse field Ising ring by disordered $ XXZ $ spin chain 
	and 
	refer to it as the sunburst quantum $XXZ$ model.
	The subsystem Hamiltonian is,
	\begin{equation}\label{eq:H_xxz}
		H_{XXZ} = \sum_{i=1}^L \sigma_i^x \sigma_{i+1}^x + 
		\sigma_i^y\sigma_{i+1}^y + \sigma_i^z \sigma_{i+1}^z + W_i 
		\sigma_i^z, 
	\end{equation}
	where $ W_i $ is an uniform random number distributed between $ [-D, 
	D] $ with $ D $ as disorder strength. The Hamiltonian shows a 
	transition from ergodic to a many-body localized (MBL) phase as the 
	strength of the disorder is increased with the transition point at $ 
	D\approx 3.6 $ \cite{PalHus2010, LuiLafFab2015, SerPapAba2015, 
	BerLak2016}.}
Notice that the MBL phase presents a situation where the nearest 
neighbor spacings are Poisson distributed like the transverse Ising chain 
despite the lack of integrability in the traditional sense. Coupling this 
with isolated qubits through identical interaction terms, as given in Eq. 
\ref{eq:subsystems}, presents another example where two subsystems showing 
Poisson distributed spacing behavior, when coupled, lead to the 
Wigner-Dyson spacings, as shown in Fig. \ref{fig:r_tilde_avg}.



\section{Spectral fluctuation: Integrable to quantum chaotic 
	transition}\label{sec:transition} {\color{black}Spectral fluctuation 
	properties such as spacing distribution have traditionally been used 
	to classify dynamical systems. For generic integrable systems, nearest 
	neighbor spacings follow Poisson distribution \cite{Berry77b} while 
	for systems with classically chaotic limits follow Wigner-Dyson 
	statistics \cite{Bohigas84}. The nearest neighbor spacing distribution 
	is between these two limits for a general non-integrable model. 
	Calculating nearest neighbor spacing distribution requires eigenvalues 
	unfolding, which is carried out numerically for most of the system. 
	The 
	sunburst quantum Ising model in the non-interacting limit is an 
	integrable system. The integrability is broken in the presence 
	of interaction terms between the Ising ring and qubits.} To follow the 
complete integrable to chaotic transition as a function of coupling 
strength $\kappa $, we utilize the average ratio of the nearest neighbor 
spacing, which captures the statistical correlation in the same fashion as 
spacing distribution with the added advantage of relaxing the requirement 
of unfolding of spectrum \cite{OgaHus2007, AtaBogGirRou2013}. 
The ratio of nearest-neighbour spacing $\tilde{r}_n$ is defined as
\begin{equation}
	\Tilde{r}_n=\frac{\min(s_n,s_{n-1})}{\max(s_n,s_{n-1})}; \quad s_n = 
	E_{n+1}-E_n
\end{equation}
where $ E_n $ is the $ n $th eigenvalue. The ensemble-averaged ratio of 
spacing $\langle \Tilde{r} \rangle$ takes the approximate value of 0.38 
for the Poisson spectrum ({\color{black}integrable system}) while for 
Gaussian Orthogonal Ensemble is approximately 0.53 ({\color{black}quantum 
chaotic systems with orthogonal symmetry})\cite{OgaHus2007, 
AtaBogGirRou2013}.
\begin{figure}[h]
	\centering
	\includegraphics[width=\linewidth]{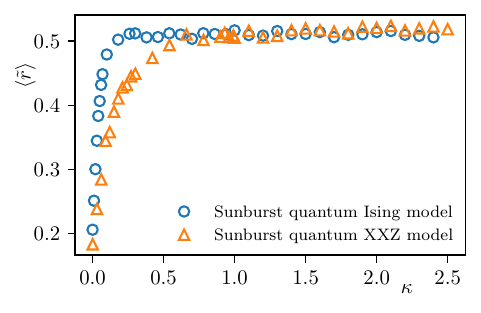}
	\caption{We plot here $\langle \Tilde{r} \rangle$ as a function 
		of 	$\kappa$. For $\kappa \lesssim 0.1$, $\langle \Tilde{r} 
		\rangle$ is closer to the Poisson limit, which indicates the 
		system is in the integrable regime. For $0.1 \lesssim \kappa 
		\lesssim 0.5$, the system is in transition from integrable to the 
		chaotic regime. For $ 0.5 \lesssim \kappa$, $\langle \Tilde{r} 
		\rangle$ is closer to the GOE limit, which indicates the system is 
		in a chaotic regime. The system parameters are chosen as $ 
		\delta=1 $, {\color{black}to break the translational symmetry 
		$h_i$s are taken as a uniformly distributed random number between 
		0.8 to 1, i.e.}  $ h_i \in \text{Unif}[0.8,1) $ along with $ L = 
		9, n=3 $. For the $ XXZ $ chain $ W_i \in \text{Unif}[-4, 4] $. 
		The ensemble of 40 realizations is considered for both models.
	}\label{fig:r_tilde_avg}
\end{figure}
The average ratio of spacing for the sunburst quantum Ising model 
initially increases almost linearly with increasing coupling strength $ 
\kappa $  then saturates around the value 0.53 (GOE) for $\kappa \gtrsim 
0.5$, as seen in Fig. \ref{fig:r_tilde_avg} (blue circles). Interestingly, 
for smaller coupling $ \kappa $, $ \langle \tilde{r} \rangle$ goes below 
Poisson value displaying the presence of Shnirelman's peak \cite{Shn1975, 
	Shn1993, ChiShe1995}. {\color{black} Let us recall that according to 
	Shnirelman's theorem, for classically nearly integrable systems, at 
	least every second spacing becomes exponentially small, causing a peak 
	in spacing distribution at zero spacing \cite{Shn1975, Shn1993}. For 
	the systems with no classical analog, this effect can be attributed to 
	discrete symmetries, which generate the quasi-degenerate pair of 
	eigenvalues. For very small $\kappa$, in the sunburst quantum Ising 
	model as well as in the sunburst quantum $XXZ$ model,  the discrete 
	permutation symmetry of interchanging qubits is almost a good symmetry 
	responsible for enhanced values of smaller spacings. We have 
	numerically verified that a small variation in $\delta_i$ removes this 
	permutation symmetry and causes the disappearance of enhanced peak at 
	zero spacing (figure not included).} We show only a portion of the 
	plot to clearly highlight the 
transition and plateauing. Utilizing this curve, we identify three regions 
(i) integrable ($ \kappa \approx 0$), (ii) {\color{black}transition 
regime} 
($ 0.1 \lesssim \kappa \lesssim 0.5$), and (iii) chaotic limit ($ \kappa 
\gtrsim 0.5 $). {\color{black}(For a single qubit connected with the Ising 
core, see Appendix \ref{ap:ln1}.)} For these three regions, the fate of an 
initial state turns out very different from each other. We probe the 
post-quench dynamics of an initial state using entanglement in Section 
\ref{sec:quench}.

\section{Post-quench dynamics: linear entropy}\label{sec:quench}
In this section, we seek to understand the role of coupling strength 
$ \kappa $ in generating the dynamics for the initial state, which is 
taken as product state of respective ground states of Ising ring and 
isolated qubits. Such a scenario is popularly known as out-of-equilibrium 
dynamics under sudden interaction quench. The quench protocol is,
\begin{equation}
	H = \begin{cases}
		H_I \otimes \unity_{q} + \unity_{I} \otimes H_q & t<0\\
		H_I \otimes \unity_{q} + \unity_{I} \otimes H_q + V_{Iq} & t\geq 
		0.
	\end{cases}
\end{equation}
We use the linear entropy of the subsystem (qubits) as a probe to 
understand the out-of-equilibrium dynamics. Linear entropy of qubits is 
defined as
\begin{equation}
	(S_L)_q= 1-\tr(\rho_q^2), 
\end{equation}
where $\rho_q$ is the reduced density matrix of the qubits obtained by 
taking a partial trace over Ising degree of freedom. A note of caution 
here is that linear entropy is not a true measure of entanglement; rather, 
it measures the amount of mixedness of the subsystems, which increases 
with increasing the entanglement between the subsystems. As a result, 
linear entropy can roughly quantify the degree of entanglement between the 
subsystems \cite{Bandyopadhyay_2005}. In the limiting case of subsystems 
being in a pure state, linear entropy is zero. As we seek to characterize 
the nature of equilibration in the large time limit, we will focus on the 
time average and variance of linear entropy.
\subsection{Limiting case: $h=0, \rev{L > 1}, n=1$}
To gain analytical insight, we derive an exact expression for 
linear entropy in the limiting case when we connect one qubit with 
the Ising ring and take {\color{black}$ h=0 $. The ground state of 
transverse Ising model with $h=0$, consistent with 
		the $\mathbb{Z}_2$  symmetry, is  $ \kt{\psi^I_G} = 
		\frac{1}{\sqrt{2}}[ \kt{+++\dots+} + \kt{---\dots-}$. This 
		macroscopic 
		superposition is known as the ``cat state''\cite{Sch1935}, or GHZ 
		state\cite{GHZ2007} in literature, and their 
generation has attracted much attention \cite{Zeilingeretal1999, 
WanBaySouSop2010, Pengetal2019, li2022rapid}. For non-zero but small $h$, 
this state is the true ground state of the 
transverse Ising chain. Although very fragile against ``symmetry 
breaking'' perturbation in $N\to \infty$ limit, this state is still 
interesting to study as our interaction term does not break the 
$\mathbb{Z}_2$ symmetry. The ground 
state of the pre-quench Hamiltonian is,
	\begin{equation}\label{eq:initial_state}
		\begin{aligned}
			\kt{\psi(0^-)}&= \kt{\psi^I_G} \otimes	\kt{0}, \text{ with}\\
			\kt{\psi^I_G} &= \frac{1}{\sqrt{2}}[ \kt{+++\dots+} + 
			\kt{---\dots-}] 
		\end{aligned}
	\end{equation}
	with $\sigma_x \kt{\pm} = \pm \kt{\pm}, \Sigma_z \kt{0} = \kt{0}$. 
	The superscript $I$ signifies the Ising part. In $h=0$ limit, Ising 
	Hamiltonian commutes with qubit and interaction part of the 
	Hamiltonian, and therefore the time-evolution operator can be 
	factorized and written as,
	\begin{align*}
		U=\exp(-iH_I t) \exp[-it(H_q+V_{Iq})]=U_IU_{Iq} = U_{Iq}U_I.
	\end{align*}
	As the initial state (Eq. \ref{eq:initial_state}) is an eigenstate of 
	$U_I$, its action on $\kt{\psi(0^-)}$ produces only a global phase 
	which we can ignore. We can evaluate $U_{Iq} \kt{\psi(0^-)} $ in 
	close form by noticing that $(H_q+V_{Iq})\kt{\psi(0^-)} = 
	-\frac{\delta}{2}\kt{\psi^I_G}\otimes \kt{0} - \kappa \kt{\psi^I_N} 
	\otimes\kt{1}$ and $(H_q + V_{Iq})^2 \kt{\psi(0^-)} = 
	\frac{\omega^2}{4} 
	\kt{\psi(0^-)}$ with $\omega^2 = \delta^2 + 4\kappa^2$.  The 
	evolved state at any time $t$ is,
	\begin{equation}\label{eq:coeff_psi_t}
		\begin{aligned}
			\kt{\psi(t)} &= A(t)\kt{\psi^I_G}\otimes 
			\kt{0}+B(t) \kt{\psi^I_N} \otimes\kt{1}, \\
			\kt{\psi^I_N}&= \frac{1}{\sqrt{2}}[ \kt{+++\dots+} - 
			\kt{---\dots-}]\\
			A(t) &= \cos \frac{\omega 
				t}{2}+i\frac{ \delta}{\omega}\sin \frac{\omega 
				t}{2}, 
			B(t) = \frac{2i
				\kappa}{\omega}\sin \frac{\omega 
				t}{2}.
		\end{aligned}
	\end{equation}   
	The quench protocol produces a superposition of two eigenstates of 
	pre-quench Hamiltonian that, interestingly, is already in Schmidt 
	form. 
	The reduced density matrix of the qubit is,
	\begin{equation}
		\rho_q(t) = \tr_I(\kt{\psi(t)}\br{\psi(t)}) = \lvert A(t)\rvert^2 
		\kt{0}\br{0} + \lvert B(t)\rvert^2 \kt{1}\br{1}.
	\end{equation}
	The linear entropy }is then given by,
\begin{align}\label{eq:lin_entropy_onequbit}
	S_L(t)=1-(\lvert A(t) \rvert^4 + \lvert B(t) \rvert^4)
\end{align}
where $ A(t), B(t) $ is defined in Eq. \ref{eq:coeff_psi_t}. It 
approaches to the maximum possible value when 
$|A(t)|^2=|B(t)|^2=\frac{1}{2}$. The time $t^*$ when the linear 
entropy reaches its maximum is given by
\begin{equation}\label{eq:max_entropy_time}
	t^*=\frac{2}{\omega}\cos^{-1}\left[\pm\sqrt{\frac{ 
			4\kappa^2-\delta^2}{8\kappa^2}}\right].
\end{equation}
It is clear from Eq. \ref{eq:max_entropy_time} that linear entropy 
can reach its maximum possible limit only if the interaction strength 
$\kappa$ is greater than or equal to half of the energy gap of the 
qubit i.e. $ 2\kappa \geq \delta $. If this condition is satisfied, 
$t^*$ decreases with increasing $\kappa$, implying that the linear 
entropy reaches its maximum value faster for the larger interaction 
strength which is what we expect intuitively. 

\begin{figure}[t]
	\centering
	\includegraphics[width=\linewidth]{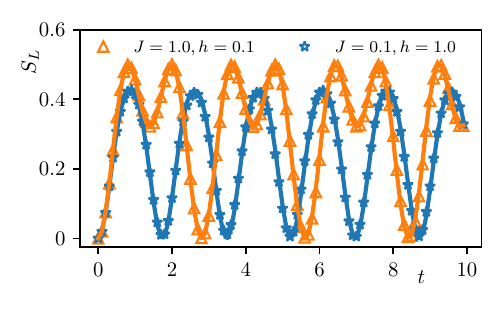}
	\caption{In two limiting cases, the analytical expression of 
		linear entropy is compared to the exact diagonalization 
		calculations. In the weak 
		field limit where $ J=1,h=0.1,\delta=\kappa=1 $ numerics is shown 
		by orange triangles while in the strong field limit where $ 	
		J=0.1,h=1,\delta=\kappa=1 $ numerics is represented by blue 
		stars. The respective analytical results (Eq. 	
		\ref{eq:lin_entropy_onequbit} with $A, B$ given by Eq. 
		\ref{eq:coeff_psi_t} and \ref{eq:coeff_psi_t_j0} respectively) are 
		plotted by solid lines. 		
		The period of the oscillation in both limits is 		
		$\frac{2\pi}{\omega}$.  
	}\label{fig:lin_entropy_large_field}
\end{figure}

\medskip
{\color{black}For small but non-zero $h$ such that $h<<J$, 
	$\kt{\psi^I_G}, 
	\kt{\psi^I_N}$ are no longer the exact eigenstate of the Ising ring, 
	however, to a very good approximation, $\kt{\psi(t)}$ continues to be 
	of the form given in Eq. \ref{eq:coeff_psi_t}. We compare the linear 
	entropy calculated in Eq. \ref{eq:lin_entropy_onequbit} with exact 
	diagonalization calculation for $h=0.1, J=1$ in Fig. 
	\ref{fig:lin_entropy_large_field}. They are in good agreement. }
\subsection{Limiting case: $J=0, \rev{L>1}, n=1$}
{\color{black}In this limit, the ground state of pre-quench 
	Hamiltonian is 
	given by, 
	\begin{equation}\label{eq:initial_state_j0}
		\begin{aligned}
			\kt{\psi(0^-)}&= \kt{\psi^I_G} \otimes	\kt{0}, \text{ with}
			\kt{\psi^I_G} &= \kt{000\dots0},
		\end{aligned}
	\end{equation}
	where $\kt{\psi^I_G} \otimes \kt{0}$ is the ground state of 
	pre-quench Hamiltonian with ground state energy $E_G = -Lh - 
	\frac{\delta}{2}$. The Ising Hamiltonian no longer commutes with 
	$V_{Iq}$ and therefore, the unitary evolution operator is given by $U 
	= \exp{-it(H_I+H_q+V_{Iq})}$. In this limit $J=0$, repeated 
	application of $H$ on the pre-quench state does not yield the 
	Identity 
	operator. We set up a difference equation for $H^n\kt{\psi(0^-)}$ in 
	terms of the initial state $\kt{\psi(0^-)}$ and state 
	$H\kt{\psi(0^-)}$ \cite{Zho2020}. Solving this difference equation 
	using 
	the 
	characteristic root method, we obtain the time-evolved state at the 
	time 
	$t$ (see Appendix \ref{ap:recursion} for the details), 
	\begin{equation}\label{eq:coeff_psi_t_j0}
		\begin{aligned}
			\kt{\psi(t)} &= A(t)\kt{\psi^I_G}\otimes 
			\kt{0}+B(t) \kt{\psi^I_N} \otimes\kt{1}, \\
			\kt{\psi^I_N}&= \kt{100\dots0}\\
			A(t) &= \cos \frac{\omega 
				t}{2}+i\frac{ \delta+2h}{\omega}\sin \frac{\omega 
				t}{2}, 
			B(t) = \frac{2i
				\kappa}{\omega}\sin \frac{\omega 
				t}{2}.
		\end{aligned}
	\end{equation}   
	Note $\kt{\psi^I_N} \otimes \kt{1}$ also is an eigenstate of 
	pre-quench Hamiltonian with eigenvalue $E_N=-(L-2) h 
	+\frac{\delta}{2}$.} The expression for linear entropy, given in 
Eq. \ref{eq:lin_entropy_onequbit} is valid in this limiting case as 
well with modified $ \omega^2=(2h+ \delta)^2+4\kappa^2 $. 

\medskip
{\color{black}The oscillatory behavior of linear entropy clearly 
	shows a lack of equilibration. Like the previous limit, for small but 
	non-zero $J<<h$, the expression of linear entropy matches very well 
	with exact diagonalization calculation as $\kt{\psi(t)}$ continues to 
	be a good approximation for finite but small $J$. The analytical 
	expression and exact diagonalization result for parameters $J=0.1, 
	h=1$ are compared in Fig. \ref{fig:lin_entropy_large_field} and are in 
	good agreement.}

\subsection{General case: $J=1, h\approx\delta \approx \kappa \approx 
	O(1), \rev{L>1, n >1}$}
For the parameter regime, $ h \approx \delta \approx \kappa \approx 
O(1) $, the total system is in the chaotic regime (see Fig. 
\ref{fig:r_tilde_avg}). An analytical solution for the time evolution 
of the linear entropy is beyond this method. To understand the time 
evolution of linear entropy under interaction quench and bring out 
the role played by \textit{coherence} of the initial state, we turn 
to exact diagonalization. In this case, linear entropy tends to 
saturate around a mean closer to the Lubkin value.
When the large time limit of linear entropy 
corresponds to the Lubkin value \cite{Lub1978}, 
we call such an equilibration thermalization.

In the rest of the paper, we try to understand perturbatively the 
initial 
growth of the linear entropy and, using this, whether or not a complete 
transition happens to a thermalized state. The emphasis will be on 
the role of initial coherence in the thermalization process characterized 
here by linear entropy. The exact diagonalization calculations for both 
models are done for system size $L=9$ and $n=1,3$ along with $J=1, 
h\approx 1$ unless mentioned otherwise.

\section{Role of initial state coherence in post-quench dynamics
}\label{sec:coherence}
The coherence of a state is a basis-dependent quantity, and we prefer 
to choose the energy basis to quantify the coherence. A state is 
called 
\textit{incoherent} in the given basis set $ \mathcal{B} = 
\{\kt{m}\}_{m=1}^N $ if the density matrix corresponding to this state is 
diagonal in the said basis. A deviation from this earns the name 
\textit{coherent}. We use the sum of the square of off-diagonal elements 
of the density matrix $ \rho $ as coherence measure \cite{BauCraPle2014},
\[ c_\mathcal{B}^2(\rho) = \sum_{m\neq m'} \lvert 
\rho_{mm^\prime}\rvert^2. 
\]
A maximally coherent state in this basis is given by,
\begin{equation}\label{eq:coh_state_defn}
	\kt{\alpha\rev{_c}}=\frac{1}{\sqrt{N}}\sum_{m=1}^N e^{i\phi_m} \kt{m}
\end{equation}
where $\phi_m\in\text{Unif}[0,2\pi)$ and can be chosen randomly 
\cite{PenJiaFan2016}. Recently, the role of coherence has been 
explored in the thermalization of the initial state using random matrix 
theory, where individual subsystems are modeled by random matrices 
\cite{Arul_2022}. It has been shown that the presence or 
lack of coherence in the initial state results in strong or weak 
thermalization.
In this section, we first obtain perturbatively the very short time 
behavior of post-quench linear entropy for the initial state chosen as (a) 
the direct product of an incoherent state (ground state of the Ising ring 
and ground state of a single qubit), (b) the direct product of maximally 
coherent state constructed using all the eigenstates of the Ising ring 
with the ground state of a single qubit. By redefining the interaction 
strength, we can write down the result for many qubits using the 
conjecture proposed in \cite{Franchi_2022}. 
Finally, we show that when time evolved by a strongly interacting 
sunburst quantum Ising model, the maximally coherent initial state 
thermalizes while the incoherent state continues to fluctuate about the 
long-time average value.
\subsection{Short-time behaviour}
Let us take the initial state of the sunburst quantum Ising ring with a 
single qubit as
\begin{equation}\label{eq:maximal_cohe_ising}
	|\psi(0^-)\rangle=  \kt{\alpha_c}	\otimes \kt{0} = 
	\frac{1}{\sqrt{2^L}}\sum_{m=1}^{2^L} e^{i\phi_m} \kt{\psi^I_m} \otimes 
	\kt{0},
\end{equation}
 with $ \kt{\psi^I_m} $ as $m$th eigenstates of the 
	Ising ring and $\phi_m\in \text{Unif}[0,2\pi)$ \rev{i.e., $\phi_m$ is 
	chosen randomly for each $m$ from uniform distribution} as in Eq. 
	\ref{eq:coh_state_defn}. In the interaction picture, the time 
	evolution operator can be approximated for a very short time 
	($t<<\kappa^{-1}$) by $	U_I(t)\approx \exp(-i V_{Iq}t) $ (for 
	details see Appendix \ref{app:evolop-intpic}). \rev{Time evolved state 
	corresponding to the initial state given in Eq. 
	\ref{eq:maximal_cohe_ising} can be calculated using the series 
	expansion of $\exp(-iV_{Iq}t)$ and re-summing the series after 
	applying each term on the initial state. For applying $V_{Iq}$ on the 
	initial state, it is sensible to expand 
	each of the eigenstates as a linear superposition of $\sigma_z$ 
	eigenbasis, i.e.
	\begin{equation}\label{eq:oneqbit_numberbasis}
		\begin{aligned}
		\kt{\alpha_c} &= \frac{1}{\sqrt{2^L}}\sum_{m=1}^{2^L} e^{i\phi_m} 
	\sum_{n=1}^{2^L}c_{nm}\kt{n}, \\ & \text{where } \kt{n} \in 
	\{\kt{00...0}, \kt{00...1}, ...\kt{11...1}\}.
	\end{aligned}
		\end{equation} 
	The action of $V_{Iq}$ once on initial state will produce, the state 
	$\kt{\alpha^\prime_c} \otimes \kt{1}$ with,
	 \begin{equation}\label{eq:oneqbit_viq}
	 	\begin{aligned}
	 		\kt{\alpha^\prime_c} &= \frac{1}{\sqrt{2^L}}\sum_{m=1}^{2^L} 
	 		e^{i\phi_m}\sum_{n=1}^{2^L}c_{nm}\kt{\tilde{n}}, \text{ where 
	 		} 
	 		\kt{\tilde{n}} = \sigma_1^x\kt{n}.
	 	\end{aligned}
	 	\end{equation}
 	Applying $V_{Iq}$ twice on the initial state will return the initial 
 	state as is evident from Eq. \ref{eq:oneqbit_numberbasis}, 
 	\ref{eq:oneqbit_viq}.} 
	This helps in obtaining a closed-form expression 
	for $ \kt{\psi(t)} $ as
	\begin{equation}
	\begin{aligned}
		|\psi(t)\rangle &=\cos(\kappa t) \kt{\alpha_c}\otimes\kt{0} + i 
		\sin(\kappa t)\kt{\alpha^\prime_c}\otimes\kt{1}.
	\end{aligned}
	\end{equation}
The subsystem linear entropy 
at time $ t $ is,
\begin{equation}
	S_L=\frac{1}{4}\left[1-\cos(4\kappa 
	t)\right]{\color{black}\left[1-\lvert 
		\gamma \rvert^2\right]}
\end{equation}
where
\begin{equation}
	\gamma=\langle \alpha_c| \alpha_c^\prime\rangle.
\end{equation}
Depending on the initial coherence, $\gamma$ changes. Let us take the 
extreme case when $ N=1 $ \textit{i.e.}  only the ground state of the 
Ising ring is taken, which is an example of \emph{incoherent} initial 
state.
Note that $ \kt{\alpha_c^\prime} $ is orthogonal to $ \kt{\alpha_c} $ 
in this case, and therefore $ \gamma = 0 $. The fact that parity 
commutes with the Ising Hamiltonian and anti-commutes with 
interaction, results in the orthogonality of $ \kt{\alpha_c} $ and $ 
\kt{\alpha_c^\prime} $.
The linear entropy for an incoherent initial state then becomes,
\begin{equation}
	S_L=\frac{1}{4}\left(1-\cos(4\kappa t)\right),
\end{equation}
which, for a very short time,
\begin{equation}\label{eq:sl_1qbit_initial}
	S_L \approx 2\kappa^2t^2.
\end{equation}
\begin{figure}[t]
	\begin{center}
		\includegraphics[width=\linewidth]{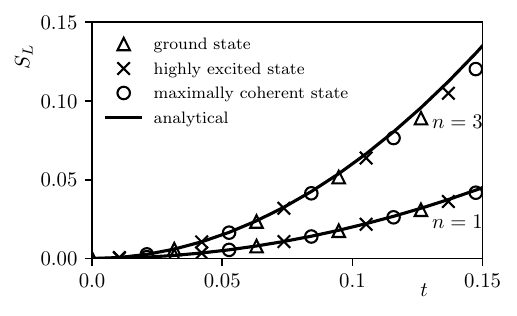}
		\caption{Initial growth of linear entropy is plotted for 
			different initial states, such as two incoherent states 
			constructed from (a) the ground state ($\triangle$) of the 
			Ising 
			ring, (b) a state from the middle of the spectrum of the Ising 
			ring ($\times$) and the maximally coherent state ($\circ$) as 
			a function of time. The length of the Ising ring is taken as 
			$L=9$ 
			for all the sets while the number of qubits is taken as $n=1$ 
			and 3. The solid line for $n=1$ is the quadratic curve derived 
			in Eq. \ref{eq:sl_1qbit_initial}, while for $n=3$ denotes the 
			Eq. \ref{eq:slnqbit}.  The other system parameters are 
			$h=0.95, \delta=1, \kappa=1$. The quadratic behavior of linear 
			entropy for a short time is clearly visible for all the 
			initial states. 	}\label{fig:coh_ini_time_decrease}
	\end{center}
\end{figure}
This quadratic dependence on time is a clear departure from earlier 
studied linear dependence of linear entropy on time 
\cite{TanFujMiy2002, BanLak2004, Cal2005,KimHus2013,KauGreetal2016}. 
{\color{black}A maximal coherent state with random phases can be 
	expanded in $\sigma_z$ eigenbasis} with random 
	coefficients \rev{as given in Eq. \ref{eq:oneqbit_numberbasis}}. The 
	action of 
	$V_{Iq}$ on this will produce another unit vector with random 
	coefficients \rev{as defined in Eq. \ref{eq:oneqbit_viq}}. \rev{For 
	maximally coherent state, we approximate $\kt{\alpha_c}, 
	\kt{\alpha^\prime_c} $ by two independent random unit vectors and as 
	$\lvert 
	\gamma \rvert$ is defined as the dot product of these two random 
	vectors, average of $\lvert \gamma \rvert^2$ over the random phases is 
	equal to the variance of the dot product of two unit random 
	vectors which is of the order of $1/2^L$ \cite{cho2009inner, 
	kus2022remark}. To 
	check the approximation 
	of taking $\kt{\alpha_c}, \kt{\alpha^\prime_c}$ as random vectors in 
	$\sigma_z$ eigenbasis, we 
	have numerically checked that average of $\lvert \gamma \rvert^2$ 
	over random phases scales as  $1/2^L$ (figure not shown here).}
	This contribution vanishes in a large $L$ limit.
Let us recall again that this behavior applies to a very short time during 
which the interaction propagator is written in terms of only $ V_{Iq} $.  

For the Ising ring coupled with two qubits, the short-time behavior 
of the linear entropy continues to be quadratic, specifically as $ 
S_L \approx 4 \kappa^2 t^2 $ (for details see Appendix 
\ref{sec:two_qubit}). This result is in agreement with the scaling 
conjecture numerically verified in \cite{Franchi_2022}.  Motivated by 
this, for the Ising ring connected with $ n $-qubits, the growth of 
linear entropy is taken as, 
\begin{equation}\label{eq:slnqbit}
	S_L \approx 2 n\kappa^2 t^2.
\end{equation}
This conjecture agrees very well with exact diagonalization 
calculations as seen in Fig. \ref{fig:coh_ini_time_decrease}  where 
the length of the Ising ring is $ L=9 $ and the number of qubits is $ 
n=3 $.

\subsection{Long time averaged entropy}
\begin{figure}
	\centering
	\includegraphics[width=\linewidth]{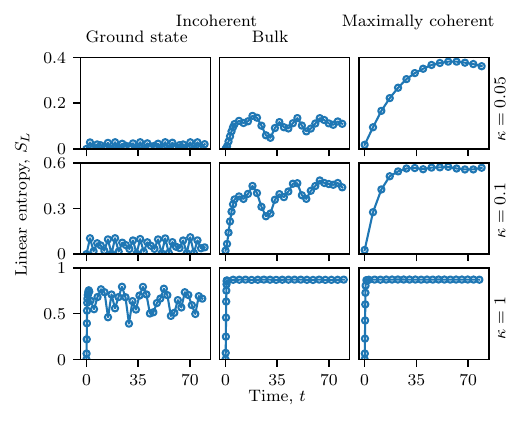}
	\caption{Time variation of linear entropy for incoherent and 
		maximally coherent initial state in ultra-weak ($ \kappa=0.05 $), 
		weak 
		($ \kappa=0.1 $) and strong interaction ($ \kappa=1 $) 
		regimes. We have considered $L=9,n=3$ and 
		$J=\delta=1, h=0.95$.}\label{fig:chaos-coh-linentropy}
\end{figure}

Not surprisingly, when we time evolve the direct product of the ground 
state of the Ising ring with the ground state of non-interacting qubits 
(\textit{i.e.} incoherent state) by near-integrable quenched Hamiltonian 
($ \kappa =0.05 $), the linear entropy fluctuates a lot near zero value 
(see Fig. \ref{fig:chaos-coh-linentropy}). For a larger value of 
interaction strength ($ \kappa =1 $) when quench Hamiltonian has a 
spectrum with Wigner-Dyson  spacing, the time-averaged linear entropy 
approaches the Lubkin value (random vector value) with visible 
fluctuations around it (see Fig. \ref{fig:chaos-coh-linentropy}). 
{\color{black} In the second column of 
	Fig. \ref{fig:chaos-coh-linentropy}, we have shown the 
	entanglement 	generation in an incoherent state which is a direct 
	product of the 
	eigenstate from the middle of the spectrum of the Ising 
	Hamiltonian 
	and the ground state of non-interacting qubits. The entanglement 
	generated by the near-integrable case is more than the incoherent 
	state corresponding to the ground state but less than the 
	maximally 
	coherent initial state. Only in the case of large $\kappa (=1)$, 
	the 
	entanglement generation is the same as in the case of a maximal
	coherent state.}

On the other hand, we have plotted linear entropy with time in the 
third column when the initial state is taken as a maximally coherent 
state and evolved in time with the quenched Hamiltonian of the same 
coupling strength as in the first and second columns of Fig. 
\ref{fig:chaos-coh-linentropy}. 
{\color{black}It is clear that the magnitude of fluctuations is 
	significantly lower 
	compared to the incoherent initial state counterparts as well as more 
	entanglement has been generated in case of $\kappa = 0.05, 0.1$. 
	Infinite time-averaged entropy and variance to capture temporal 
	fluctuation around it are defined as,
	\begin{equation}\label{eq:sl_aver_def}
		\begin{aligned}
			\langle S_L \rangle &= \lim_{\tau \to \infty}\frac{1}{\tau} 
			\int_0^\tau 
			S_L(t) dt,\\
			\sigma^2(S_L) &= \langle (S_L(t) - \langle S_L\rangle 
			)^2\rangle.
		\end{aligned}
\end{equation}}
We have tabulated the effect of coherence on the long-time averaged 
value of entropy along with variance to quantify this fluctuation in 
Tab. \ref{tab:linent_variance_coherence}.
\begin{table}[h]
	\begin{tabular}{|c|c|c|c|c|}
		\hline
		&  \multicolumn{2}{c|}{$L = 9, n=1$}   & \multicolumn{2}{c|}{$L = 
			9, n=3$} \\
		\hline
		$c_\mathcal{B}^2$ & $\langle 
		S_L(t) \rangle$ & $\sigma^2(S_L)$ & 
		$\langle 
		S_L(t) \rangle$  & 
		$\sigma^2(S_L)$ \\
		\hline
		0.5000& 0.2309 & 0.1103 & 0.5704 & 0.0651 \\
		\hline
		0.7500&  0.3236 & 0.0766 & 0.6959 & 0.0488  \\
		\hline
		0.8750& 0.3634 & 0.0635 & 0.7587 & 0.0247 \\
		\hline
		0.9375& 0.4100 & 0.0393 & 0.7849 & 0.0147 \\
		\hline
		0.9687& 0.4217 & 0.0286 & 0.8227 & 0.0082 \\
		\hline
		0.9844& 0.4651 & 0.0152 & 0.8408 & 0.0058 \\
		\hline
		0.9922& 0.4782 & 0.0090 & 0.8560  & 0.0030  \\
		\hline
		0.9960& 0.4877 & 0.0061 & 0.8654  & 0.0013  \\
		\hline
		0.9980& 0.4974 & 0.0018 & 0.8722 & 0.0005  \\
		
		\hline
	\end{tabular}
	\caption{The time-averaged entropy and variance is listed as a 
	function of initial state 
	coherence.}\label{tab:linent_variance_coherence}
\end{table}

The variance decreases almost inversely proportional to the initial 
state coherence and, at the same time, the long-time averaged linear 
entropy increases to its limit of Lubkin value. {\color{black}This shows 
	that coherence of the initial state acts as a resource for 
	entanglement 
	generation and decreasing variance signify strong thermalization.
}

\begin{figure}[!h]
	\centering
	\includegraphics[width=0.95\linewidth]{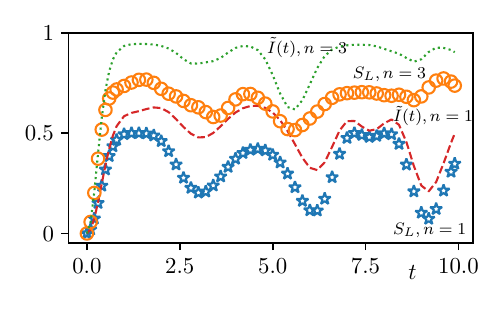}
	\caption{Time evolution of linear entropy and 
		$\tilde{I}(t)$ of the state $\kt{\psi(t)}$ evolved in time by 
		post-quench 
		sunburst quantum Ising Hamiltonian are plotted here. 
		The blue star and orange circle are linear entropy for the 
		sunburst 
		quantum Ising model for $ n=1 $ and $ n=3 $ qubits, respectively. 
		Red 
		dashed line ($ n=1 $ qubits) and green dotted line ($ n=3 $ 
		qubits) 
		are representing $\tilde{I}(t)$ for the same time evolved 
		state in pre-quench eigenbasis. For these $ \kappa = \delta= 1.0 $ 
		has been taken along with $ h = 0.95 $, and the initial state is 
		incoherent. 
		The choice of plotting $\tilde{I}(t)$ in 
		place of $I_{\kt{\psi(t)}}$ is to show the hills and valleys 
		correspond exactly to those of linear entropy.}
	\label{fig:ipr_LE}
\end{figure}

To understand the possible mechanism of fluctuations in linear 
entropy around Lubkin value seen when an incoherent state is evolved 
in time with strongly coupled quenched Hamiltonian, 
we plot the inverse participation ratio (IPR) of the time-evolved 
state in the pre-quench eigenenergy basis. {\color{black}We define IPR as,
	\begin{equation}
		\begin{aligned}
			I_{\kt{\psi(t)}} &= \sum_{m=1}^{N} \lvert \langle 
			\psi^{H_0}_m|\psi(t)\rangle\rvert^4, \quad \text{with}\\
			(H_I \otimes 
			\unity_{q} &+ \unity_{I} \otimes H_q) \kt{\psi^{H_0}_m} = E_m 
			\kt{\psi^{H_0}_m}, m=1\dots N.
		\end{aligned}
\end{equation}}
This quantity takes two extreme values: $1$ and $1/N$ (inverse of Hilbert 
space dimension) for the limits when 
the time-evolved state is 
built upon only one of the pre-quench states, and when it is built upon 
all the 
pre-quench eigenstates, respectively. To show the correlation, we 
have plotted {\color{black}$ \tilde{I}(t) \equiv 1 - 
	I_{\kt{\psi(t)}} $} (red dashed line for $ n=1 $ qubit, and green 
dotted lines for $ n=3 $ qubits) along with linear entropy (blue star 
for $ n=1 $ qubit and orange circle for $ n=3 $ qubits) as seen in 
Fig. \ref{fig:ipr_LE}. For $ n=1 $ qubit, $ \tilde{I}(t) $ 
is fluctuating and averaging to nearly half. This shows that the 
time-evolving state explores a very small subset of Hilbert space of the 
pre-quench system, and therefore we see a large fluctuation in entropy. 
This fact is borne out in the Fig. \ref{fig:ipr_LE} very clearly where the 
smaller value of IPR (i.e., a larger value of $\tilde{I}(t)$ as plotted in 
the figure) corresponds to larger entanglement entropy. {\color{black}This 
clearly shows that an increase (decrease) in entanglement entropy 
corresponds to instantaneous larger (smaller) participation of pre-quench 
eigenstates in the time-evolved incoherent initial state.} 
For the maximally coherent initial state (Eq. \ref{eq:maximal_cohe_ising}) 
on the other hand, IPR calculated in the pre-quench eigenenergy 
basis is approximately independent of time. This observation is 
intuitively clear as a maximally coherent state is constructed with all 
eigenstates of the Ising ring. The IPR comes out to be $ \approx 1/2^{L+n} 
$, which 
shows that the complete Hilbert space of the pre-quench Hamiltonian 
is explored by the time-evolving state for all time. 

\begin{figure}[t]
	\centering
	\includegraphics[width=\linewidth]{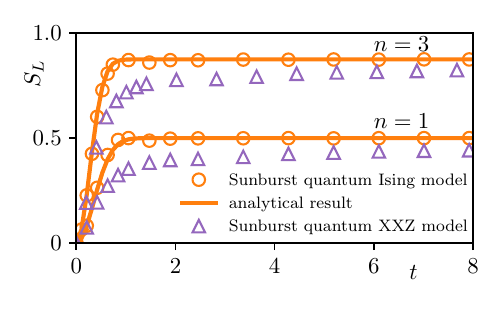}
	\caption{Evolution of linear entropy when maximally coherent product 
	state is evolved by the sunburst quantum Ising model and sunburst 
	quantum $XXZ$ model. In the $XXZ $ model, parameters are chosen in 
	such a way that the system is in the MBL phase. The solid 
		lines are analytical results derived in Eqs. \ref{eq:sl_complete}, 
		\ref{eq:sl_complete_nqbit}. The sunburst quantum $XXZ$ model where 
		qubits are connected with the ring of $ XXZ $ chain with disorder 
		strength $ D = 4 $  puts the $XXZ$ ring in the many-body localized 
		phase. The other parameters are $ \delta=1 $, $ L=9 $, and $ 
		\kappa=1.5 $.
	}\label{fig:sl_1qbit_full}
\end{figure}

For the case when complete Hilbert space is being explored, we 
derived the initial dependence of linear entropy on time to be 
quadratic (Eq. \ref{eq:sl_1qbit_initial}). Using the perturbation 
theory developed in \cite{Arul_2016} and initial time dependence of 
entropy, the complete transition of linear entropy from 0 to Lubkin 
value can be obtained (see Appendix \ref{ap:sl_deri}),
\begin{equation}\label{eq:sl_complete}
	S_L(t) = \left[1 - 
	\exp\left(-\frac{2\kappa^2t^2}{S_L^\infty}\right)\right]S_L^\infty.
\end{equation}
This form is compared with linear entropy calculated numerically for 
sunburst {\color{black} quantum Ising} model when a single qubit is 
connected with the Ising ring 
in Fig. \ref{fig:sl_1qbit_full} (orange circles).
Utilizing the initial time-dependence of entropy for the sunburst 
quantum Ising model with $ n- $qubits, Eq. \ref{eq:sl_complete} is 
generalized to, 
\begin{equation}\label{eq:sl_complete_nqbit}
	S_L(t) = \left[1 - 
	\exp\left(-\frac{2n\kappa^2t^2}{S_L^\infty}\right)\right]S_L^\infty.
\end{equation}
This result is in clear agreement with the exact diagonalization 
calculation as seen in Fig. \ref{fig:sl_1qbit_full} where $ n=3 $ qubits 
have been 
chosen (orange circles), and like in the single qubit case, the initial 
condition was a maximally coherent state.

For both cases, whether the initial state is incoherent or 
maximally coherent, the reduced density matrix corresponding to the 
qubit subspace becomes nearly diagonal, a telltale 
sign of equilibration. The fluctuation in linear entropy then captures 
whether equilibration is strong or weak. Therefore, the variance of the 
time series of linear entropy becomes a good measure to identify the 
nature of the equilibration of the time-evolved state. As seen in Table. 
\ref{tab:linent_variance_coherence}, for a fixed coupling 
strength where the total system is quantum chaotic, and variance is 
(almost) inversely proportional to the degree of coherence of the 
initial state. The time-averaged value of linear entropy becoming 
equal to the Lubkin value with a very small variance 
indicates the strong thermalization when the initial state is 
maximally coherent and time-evolution is done using post-quench 
Hamiltonian showing Wigner-Dyson statistics for its spectral 
fluctuation. On the other hand, thermalization remains weak for an 
incoherent initial state for the same value of coupling constant $ 
\kappa $.

For the sunburst quantum $XXZ$ model, the average ratio of spacing with 
coupling strength $ \kappa $ is plotted in Fig. \ref{fig:r_tilde_avg} 
(orange triangle), showing a transition from integrability to chaos. We 
choose a sufficiently large $ \kappa $ value to correspond to the average 
ratio of the spacings to correspond to Wigner-Dyson statistics. The time 
evolution of linear entropy of qubits post interaction quench in the 
sunburst quantum XXZ model is plotted in Fig. \ref{fig:sl_1qbit_full}. The 
numerical linear entropy evolution closely follows the analytical form 
derived for the sunburst quantum Ising model, as is borne out clearly from 
Fig. \ref{fig:sl_1qbit_full}. A systematic lower value of linear entropy 
from the Lubkin value may be specific to MBL physics, which we will take 
up for a future study.

\section{Summary and outlook}\label{sec:summary}
In this paper, we have studied the non-equilibrium dynamics of an isolated 
bipartite quantum system, the sunburst quantum Ising model, under 
interaction quench where pre-quench subsystems are integrable. The 
interaction strength drives the post-quench system to a quantum chaotic 
regime, with the average ratio of spacings being $ \approx 0.53 $ for $ 
\kappa \gtrsim 0.5 $ consistent with GOE of random matrix theory. We have 
derived a quadratic in {\color{black}time} growth of linear entropy for 
short time.
The coherence of the initial state is shown to slow it down a little, but 
the contribution is vanishingly small as we increase the length of the 
Ising ring. The initial state, which we chose as a product state of 
eigenstates of pre-quench Hamiltonian subsystems, equilibrates in the 
large time limit when the post-quench Hamiltonian generates the time 
evolution with interaction strength value one or larger as long as level 
spacing distribution remains Wigner-Dyson. We derived the full transition 
curve for linear entropy, which agrees with numerical calculation.
\rev{We further highlight the increase in time-averaged entropy (Eq. 
	\ref{eq:sl_aver_def}) tabulated in Tab. 1 as a function of the initial 
	state 
	coherence, suggestive of later being a resource of entanglement. The 
	decreasing variance of linear entropy with initial state coherence 
	characterizes the smallness of oscillations around the mean value, or 
	in other words, initial state coherence helps in achieving the steady 
	state behavior even in a weak interaction regime (see the second row 
	of Fig. 
	\ref{fig:chaos-coh-linentropy}).}
We have shown that the effect of coherence on the nature of equilibrium is 
a ``generic'' feature of bipartite near-integrable/quantum chaotic 
systems with individual parts taken as integrable by replacing the Ising 
ring with $ XXZ $ chain taken in the MBL phase. Though quantifying this 
effect analytically has {\color{black} earlier} been done using random 
matrix theory in the limit when pre-quench subsystems are chaotic 
\cite{Arul_2022}, it remains an open problem for integrable subsystems as 
studied here. The lack of sunburst quantum $XXZ$ model (in MBL phase) in 
attaining the Lubkin value for linear entropy despite the overall system 
showing Wigner-Dyson spacing also remains an interesting open question to 
explore.

\bibliography{references}

\begin{thebibliography}{78}%
\makeatletter
\providecommand \@ifxundefined [1]{%
 \@ifx{#1\undefined}
}%
\providecommand \@ifnum [1]{%
 \ifnum #1\expandafter \@firstoftwo
 \else \expandafter \@secondoftwo
 \fi
}%
\providecommand \@ifx [1]{%
 \ifx #1\expandafter \@firstoftwo
 \else \expandafter \@secondoftwo
 \fi
}%
\providecommand \natexlab [1]{#1}%
\providecommand \enquote  [1]{``#1''}%
\providecommand \bibnamefont  [1]{#1}%
\providecommand \bibfnamefont [1]{#1}%
\providecommand \citenamefont [1]{#1}%
\providecommand \href@noop [0]{\@secondoftwo}%
\providecommand \href [0]{\begingroup \@sanitize@url \@href}%
\providecommand \@href[1]{\@@startlink{#1}\@@href}%
\providecommand \@@href[1]{\endgroup#1\@@endlink}%
\providecommand \@sanitize@url [0]{\catcode `\\12\catcode `\$12\catcode
  `\&12\catcode `\#12\catcode `\^12\catcode `\_12\catcode `\%12\relax}%
\providecommand \@@startlink[1]{}%
\providecommand \@@endlink[0]{}%
\providecommand \url  [0]{\begingroup\@sanitize@url \@url }%
\providecommand \@url [1]{\endgroup\@href {#1}{\urlprefix }}%
\providecommand \urlprefix  [0]{URL }%
\providecommand \Eprint [0]{\href }%
\providecommand \doibase [0]{https://doi.org/}%
\providecommand \selectlanguage [0]{\@gobble}%
\providecommand \bibinfo  [0]{\@secondoftwo}%
\providecommand \bibfield  [0]{\@secondoftwo}%
\providecommand \translation [1]{[#1]}%
\providecommand \BibitemOpen [0]{}%
\providecommand \bibitemStop [0]{}%
\providecommand \bibitemNoStop [0]{.\EOS\space}%
\providecommand \EOS [0]{\spacefactor3000\relax}%
\providecommand \BibitemShut  [1]{\csname bibitem#1\endcsname}%
\let\auto@bib@innerbib\@empty
\bibitem [{\citenamefont {Kinoshita}\ \emph {et~al.}(2006)\citenamefont
  {Kinoshita}, \citenamefont {Wenger},\ and\ \citenamefont
  {Weiss}}]{Kitowe_2006}%
  \BibitemOpen
  \bibfield  {author} {\bibinfo {author} {\bibfnamefont {T.}~\bibnamefont
  {Kinoshita}}, \bibinfo {author} {\bibfnamefont {T.}~\bibnamefont {Wenger}},\
  and\ \bibinfo {author} {\bibfnamefont {D.}~\bibnamefont {Weiss}},\ }\bibfield
   {title} {\bibinfo {title} {A quantum newton's cradle},\ }\href
  {https://doi.org/10.1038/nature04693} {\bibfield  {journal} {\bibinfo
  {journal} {Nature}\ }\textbf {\bibinfo {volume} {440}},\ \bibinfo {pages}
  {900} (\bibinfo {year} {2006})}\BibitemShut {NoStop}%
\bibitem [{\citenamefont {Cheneau}\ \emph {et~al.}(2012)\citenamefont
  {Cheneau}, \citenamefont {Barmettler}, \citenamefont {Poletti}, \citenamefont
  {Endres}, \citenamefont {Schau{\ss}}, \citenamefont {Fukuhara}, \citenamefont
  {Gross}, \citenamefont {Bloch}, \citenamefont {Kollath},\ and\ \citenamefont
  {Kuhr}}]{Cheneau_2012}%
  \BibitemOpen
  \bibfield  {author} {\bibinfo {author} {\bibfnamefont {M.}~\bibnamefont
  {Cheneau}}, \bibinfo {author} {\bibfnamefont {P.}~\bibnamefont {Barmettler}},
  \bibinfo {author} {\bibfnamefont {D.}~\bibnamefont {Poletti}}, \bibinfo
  {author} {\bibfnamefont {M.}~\bibnamefont {Endres}}, \bibinfo {author}
  {\bibfnamefont {P.}~\bibnamefont {Schau{\ss}}}, \bibinfo {author}
  {\bibfnamefont {T.}~\bibnamefont {Fukuhara}}, \bibinfo {author}
  {\bibfnamefont {C.}~\bibnamefont {Gross}}, \bibinfo {author} {\bibfnamefont
  {I.}~\bibnamefont {Bloch}}, \bibinfo {author} {\bibfnamefont
  {C.}~\bibnamefont {Kollath}},\ and\ \bibinfo {author} {\bibfnamefont
  {S.}~\bibnamefont {Kuhr}},\ }\bibfield  {title} {\bibinfo {title}
  {Light-cone-like spreading of correlations in a quantum many-body system},\
  }\href {https://doi.org/10.1038/nature10748} {\bibfield  {journal} {\bibinfo
  {journal} {Nature}\ }\textbf {\bibinfo {volume} {481}},\ \bibinfo {pages}
  {484} (\bibinfo {year} {2012})}\BibitemShut {NoStop}%
\bibitem [{\citenamefont {Langen}\ \emph {et~al.}(2015)\citenamefont {Langen},
  \citenamefont {Erne}, \citenamefont {Geiger}, \citenamefont {Rauer},
  \citenamefont {Schweigler}, \citenamefont {Kuhnert}, \citenamefont
  {Rohringer}, \citenamefont {Mazets}, \citenamefont {Gasenzer},\ and\
  \citenamefont {Schmiedmayer}}]{Langen_2015}%
  \BibitemOpen
  \bibfield  {author} {\bibinfo {author} {\bibfnamefont {T.}~\bibnamefont
  {Langen}}, \bibinfo {author} {\bibfnamefont {S.}~\bibnamefont {Erne}},
  \bibinfo {author} {\bibfnamefont {R.}~\bibnamefont {Geiger}}, \bibinfo
  {author} {\bibfnamefont {B.}~\bibnamefont {Rauer}}, \bibinfo {author}
  {\bibfnamefont {T.}~\bibnamefont {Schweigler}}, \bibinfo {author}
  {\bibfnamefont {M.}~\bibnamefont {Kuhnert}}, \bibinfo {author} {\bibfnamefont
  {W.}~\bibnamefont {Rohringer}}, \bibinfo {author} {\bibfnamefont {I.~E.}\
  \bibnamefont {Mazets}}, \bibinfo {author} {\bibfnamefont {T.}~\bibnamefont
  {Gasenzer}},\ and\ \bibinfo {author} {\bibfnamefont {J.}~\bibnamefont
  {Schmiedmayer}},\ }\bibfield  {title} {\bibinfo {title} {Experimental
  observation of a generalized gibbs ensemble},\ }\href
  {https://doi.org/10.1126/science.1257026} {\bibfield  {journal} {\bibinfo
  {journal} {Science}\ }\textbf {\bibinfo {volume} {348}},\ \bibinfo {pages}
  {207} (\bibinfo {year} {2015})}\BibitemShut {NoStop}%
\bibitem [{\citenamefont {Brydges}\ \emph {et~al.}(2019)\citenamefont
  {Brydges}, \citenamefont {Elben}, \citenamefont {Jurcevic}, \citenamefont
  {Vermersch}, \citenamefont {Maier}, \citenamefont {Lanyon}, \citenamefont
  {Zoller}, \citenamefont {Blatt},\ and\ \citenamefont {Roos}}]{Brydges_2019}%
  \BibitemOpen
  \bibfield  {author} {\bibinfo {author} {\bibfnamefont {T.}~\bibnamefont
  {Brydges}}, \bibinfo {author} {\bibfnamefont {A.}~\bibnamefont {Elben}},
  \bibinfo {author} {\bibfnamefont {P.}~\bibnamefont {Jurcevic}}, \bibinfo
  {author} {\bibfnamefont {B.}~\bibnamefont {Vermersch}}, \bibinfo {author}
  {\bibfnamefont {C.}~\bibnamefont {Maier}}, \bibinfo {author} {\bibfnamefont
  {B.~P.}\ \bibnamefont {Lanyon}}, \bibinfo {author} {\bibfnamefont
  {P.}~\bibnamefont {Zoller}}, \bibinfo {author} {\bibfnamefont
  {R.}~\bibnamefont {Blatt}},\ and\ \bibinfo {author} {\bibfnamefont {C.~F.}\
  \bibnamefont {Roos}},\ }\bibfield  {title} {\bibinfo {title} {Probing
  r{\'{e}}nyi entanglement entropy via randomized measurements},\ }\href
  {https://doi.org/10.1126/science.aau4963} {\bibfield  {journal} {\bibinfo
  {journal} {Science}\ }\textbf {\bibinfo {volume} {364}},\ \bibinfo {pages}
  {260} (\bibinfo {year} {2019})}\BibitemShut {NoStop}%
\bibitem [{\citenamefont {Lukin}\ \emph {et~al.}(2019)\citenamefont {Lukin},
  \citenamefont {Rispoli}, \citenamefont {Schittko}, \citenamefont {Tai},
  \citenamefont {Kaufman}, \citenamefont {Choi}, \citenamefont {Khemani},
  \citenamefont {L{\'{e} }onard},\ and\ \citenamefont {Greiner}}]{Lukin_2019}%
  \BibitemOpen
  \bibfield  {author} {\bibinfo {author} {\bibfnamefont {A.}~\bibnamefont
  {Lukin}}, \bibinfo {author} {\bibfnamefont {M.}~\bibnamefont {Rispoli}},
  \bibinfo {author} {\bibfnamefont {R.}~\bibnamefont {Schittko}}, \bibinfo
  {author} {\bibfnamefont {M.~E.}\ \bibnamefont {Tai}}, \bibinfo {author}
  {\bibfnamefont {A.~M.}\ \bibnamefont {Kaufman}}, \bibinfo {author}
  {\bibfnamefont {S.}~\bibnamefont {Choi}}, \bibinfo {author} {\bibfnamefont
  {V.}~\bibnamefont {Khemani}}, \bibinfo {author} {\bibfnamefont
  {J.}~\bibnamefont {L{\'{e} }onard}},\ and\ \bibinfo {author} {\bibfnamefont
  {M.}~\bibnamefont {Greiner}},\ }\bibfield  {title} {\bibinfo {title} {Probing
  entanglement in a many-body{\textendash}localized system},\ }\href
  {https://doi.org/10.1126/science.aau0818} {\bibfield  {journal} {\bibinfo
  {journal} {Science}\ }\textbf {\bibinfo {volume} {364}},\ \bibinfo {pages}
  {256} (\bibinfo {year} {2019})}\BibitemShut {NoStop}%
\bibitem [{\citenamefont {Schneider}\ \emph {et~al.}(2012)\citenamefont
  {Schneider}, \citenamefont {Hackermüller}, \citenamefont {Ronzheimer},
  \citenamefont {Will}, \citenamefont {Braun}, \citenamefont {Best},
  \citenamefont {Bloch}, \citenamefont {Demler}, \citenamefont {Mandt},
  \citenamefont {Rasch},\ and\ \citenamefont {Rosch}}]{Schneider_2012}%
  \BibitemOpen
  \bibfield  {author} {\bibinfo {author} {\bibfnamefont {U.}~\bibnamefont
  {Schneider}}, \bibinfo {author} {\bibfnamefont {L.}~\bibnamefont
  {Hackermüller}}, \bibinfo {author} {\bibfnamefont {J.~P.}\ \bibnamefont
  {Ronzheimer}}, \bibinfo {author} {\bibfnamefont {S.}~\bibnamefont {Will}},
  \bibinfo {author} {\bibfnamefont {S.}~\bibnamefont {Braun}}, \bibinfo
  {author} {\bibfnamefont {T.}~\bibnamefont {Best}}, \bibinfo {author}
  {\bibfnamefont {I.}~\bibnamefont {Bloch}}, \bibinfo {author} {\bibfnamefont
  {E.}~\bibnamefont {Demler}}, \bibinfo {author} {\bibfnamefont
  {S.}~\bibnamefont {Mandt}}, \bibinfo {author} {\bibfnamefont
  {D.}~\bibnamefont {Rasch}},\ and\ \bibinfo {author} {\bibfnamefont
  {A.}~\bibnamefont {Rosch}},\ }\bibfield  {title} {\bibinfo {title} {Fermionic
  transport and out-of-equilibrium dynamics in a homogeneous hubbard
  model~with~ultracold atoms},\ }\href {https://doi.org/10.1038/nphys2205}
  {\bibfield  {journal} {\bibinfo  {journal} {Nature Physics}\ }\textbf
  {\bibinfo {volume} {8}},\ \bibinfo {pages} {213} (\bibinfo {year}
  {2012})}\BibitemShut {NoStop}%
\bibitem [{\citenamefont {Gring}\ \emph {et~al.}(2012)\citenamefont {Gring},
  \citenamefont {Kuhnert}, \citenamefont {Langen}, \citenamefont {Kitagawa},
  \citenamefont {Rauer}, \citenamefont {Schreitl}, \citenamefont {Mazets},
  \citenamefont {Smith}, \citenamefont {Demler},\ and\ \citenamefont
  {Schmiedmayer}}]{Gring_2012}%
  \BibitemOpen
  \bibfield  {author} {\bibinfo {author} {\bibfnamefont {M.}~\bibnamefont
  {Gring}}, \bibinfo {author} {\bibfnamefont {M.}~\bibnamefont {Kuhnert}},
  \bibinfo {author} {\bibfnamefont {T.}~\bibnamefont {Langen}}, \bibinfo
  {author} {\bibfnamefont {T.}~\bibnamefont {Kitagawa}}, \bibinfo {author}
  {\bibfnamefont {B.}~\bibnamefont {Rauer}}, \bibinfo {author} {\bibfnamefont
  {M.}~\bibnamefont {Schreitl}}, \bibinfo {author} {\bibfnamefont
  {I.}~\bibnamefont {Mazets}}, \bibinfo {author} {\bibfnamefont {D.~A.}\
  \bibnamefont {Smith}}, \bibinfo {author} {\bibfnamefont {E.}~\bibnamefont
  {Demler}},\ and\ \bibinfo {author} {\bibfnamefont {J.}~\bibnamefont
  {Schmiedmayer}},\ }\bibfield  {title} {\bibinfo {title} {Relaxation and
  prethermalization in an isolated quantum system},\ }\href
  {https://doi.org/10.1126/science.1224953} {\bibfield  {journal} {\bibinfo
  {journal} {Science}\ }\textbf {\bibinfo {volume} {337}},\ \bibinfo {pages}
  {1318} (\bibinfo {year} {2012})}\BibitemShut {NoStop}%
\bibitem [{\citenamefont {Kaufman}\ \emph
  {et~al.}(2016{\natexlab{a}})\citenamefont {Kaufman}, \citenamefont {Tai},
  \citenamefont {Lukin}, \citenamefont {Rispoli}, \citenamefont {Schittko},
  \citenamefont {Preiss},\ and\ \citenamefont {Greiner}}]{Kaufman_2016}%
  \BibitemOpen
  \bibfield  {author} {\bibinfo {author} {\bibfnamefont {A.~M.}\ \bibnamefont
  {Kaufman}}, \bibinfo {author} {\bibfnamefont {M.~E.}\ \bibnamefont {Tai}},
  \bibinfo {author} {\bibfnamefont {A.}~\bibnamefont {Lukin}}, \bibinfo
  {author} {\bibfnamefont {M.}~\bibnamefont {Rispoli}}, \bibinfo {author}
  {\bibfnamefont {R.}~\bibnamefont {Schittko}}, \bibinfo {author}
  {\bibfnamefont {P.~M.}\ \bibnamefont {Preiss}},\ and\ \bibinfo {author}
  {\bibfnamefont {M.}~\bibnamefont {Greiner}},\ }\bibfield  {title} {\bibinfo
  {title} {Quantum thermalization through entanglement in an isolated many-body
  system},\ }\href {https://doi.org/10.1126/science.aaf6725} {\bibfield
  {journal} {\bibinfo  {journal} {Science}\ }\textbf {\bibinfo {volume}
  {353}},\ \bibinfo {pages} {794} (\bibinfo {year}
  {2016}{\natexlab{a}})}\BibitemShut {NoStop}%
\bibitem [{\citenamefont {Trotzky}\ \emph {et~al.}(2012)\citenamefont
  {Trotzky}, \citenamefont {Chen}, \citenamefont {Flesch}, \citenamefont
  {McCulloch}, \citenamefont {Schollwöck}, \citenamefont {Eisert},\ and\
  \citenamefont {Bloch}}]{Trotzky_2012}%
  \BibitemOpen
  \bibfield  {author} {\bibinfo {author} {\bibfnamefont {S.}~\bibnamefont
  {Trotzky}}, \bibinfo {author} {\bibfnamefont {Y.-A.}\ \bibnamefont {Chen}},
  \bibinfo {author} {\bibfnamefont {A.}~\bibnamefont {Flesch}}, \bibinfo
  {author} {\bibfnamefont {I.~P.}\ \bibnamefont {McCulloch}}, \bibinfo {author}
  {\bibfnamefont {U.}~\bibnamefont {Schollwöck}}, \bibinfo {author}
  {\bibfnamefont {J.}~\bibnamefont {Eisert}},\ and\ \bibinfo {author}
  {\bibfnamefont {I.}~\bibnamefont {Bloch}},\ }\bibfield  {title} {\bibinfo
  {title} {Probing the relaxation towards equilibrium in an isolated strongly
  correlated one-dimensional bose~gas},\ }\href
  {https://doi.org/10.1038/nphys2232} {\bibfield  {journal} {\bibinfo
  {journal} {Nature Physics}\ }\textbf {\bibinfo {volume} {8}},\ \bibinfo
  {pages} {325} (\bibinfo {year} {2012})}\BibitemShut {NoStop}%
\bibitem [{\citenamefont {Langen}\ \emph {et~al.}(2013)\citenamefont {Langen},
  \citenamefont {Geiger}, \citenamefont {Kuhnert}, \citenamefont {Rauer},\ and\
  \citenamefont {Schmiedmayer}}]{Langen_2013}%
  \BibitemOpen
  \bibfield  {author} {\bibinfo {author} {\bibfnamefont {T.}~\bibnamefont
  {Langen}}, \bibinfo {author} {\bibfnamefont {R.}~\bibnamefont {Geiger}},
  \bibinfo {author} {\bibfnamefont {M.}~\bibnamefont {Kuhnert}}, \bibinfo
  {author} {\bibfnamefont {B.}~\bibnamefont {Rauer}},\ and\ \bibinfo {author}
  {\bibfnamefont {J.}~\bibnamefont {Schmiedmayer}},\ }\bibfield  {title}
  {\bibinfo {title} {Local emergence of thermal correlations in an isolated
  quantum many-body system},\ }\href {https://doi.org/10.1038/nphys2739}
  {\bibfield  {journal} {\bibinfo  {journal} {Nature Physics}\ }\textbf
  {\bibinfo {volume} {9}},\ \bibinfo {pages} {640} (\bibinfo {year}
  {2013})}\BibitemShut {NoStop}%
\bibitem [{\citenamefont {Islam}\ \emph {et~al.}(2015)\citenamefont {Islam},
  \citenamefont {Ma}, \citenamefont {Preiss}, \citenamefont {Eric~Tai},
  \citenamefont {Lukin}, \citenamefont {Rispoli},\ and\ \citenamefont
  {Greiner}}]{Islam2015}%
  \BibitemOpen
  \bibfield  {author} {\bibinfo {author} {\bibfnamefont {R.}~\bibnamefont
  {Islam}}, \bibinfo {author} {\bibfnamefont {R.}~\bibnamefont {Ma}}, \bibinfo
  {author} {\bibfnamefont {P.~M.}\ \bibnamefont {Preiss}}, \bibinfo {author}
  {\bibfnamefont {M.}~\bibnamefont {Eric~Tai}}, \bibinfo {author}
  {\bibfnamefont {A.}~\bibnamefont {Lukin}}, \bibinfo {author} {\bibfnamefont
  {M.}~\bibnamefont {Rispoli}},\ and\ \bibinfo {author} {\bibfnamefont
  {M.}~\bibnamefont {Greiner}},\ }\bibfield  {title} {\bibinfo {title}
  {Measuring entanglement entropy in a quantum many-body system},\ }\href
  {https://doi.org/10.1038/nature15750} {\bibfield  {journal} {\bibinfo
  {journal} {Nature}\ }\textbf {\bibinfo {volume} {528}},\ \bibinfo {pages}
  {77} (\bibinfo {year} {2015})}\BibitemShut {NoStop}%
\bibitem [{\citenamefont {Ritter}\ \emph {et~al.}(2007)\citenamefont {Ritter},
  \citenamefont {\"Ottl}, \citenamefont {Donner}, \citenamefont {Bourdel},
  \citenamefont {K\"ohl},\ and\ \citenamefont {Esslinger}}]{Ritter_2007}%
  \BibitemOpen
  \bibfield  {author} {\bibinfo {author} {\bibfnamefont {S.}~\bibnamefont
  {Ritter}}, \bibinfo {author} {\bibfnamefont {A.}~\bibnamefont {\"Ottl}},
  \bibinfo {author} {\bibfnamefont {T.}~\bibnamefont {Donner}}, \bibinfo
  {author} {\bibfnamefont {T.}~\bibnamefont {Bourdel}}, \bibinfo {author}
  {\bibfnamefont {M.}~\bibnamefont {K\"ohl}},\ and\ \bibinfo {author}
  {\bibfnamefont {T.}~\bibnamefont {Esslinger}},\ }\bibfield  {title} {\bibinfo
  {title} {Observing the formation of long-range order during bose-einstein
  condensation},\ }\href {https://doi.org/10.1103/PhysRevLett.98.090402}
  {\bibfield  {journal} {\bibinfo  {journal} {Phys. Rev. Lett.}\ }\textbf
  {\bibinfo {volume} {98}},\ \bibinfo {pages} {090402} (\bibinfo {year}
  {2007})}\BibitemShut {NoStop}%
\bibitem [{\citenamefont {Deutsch}(1991)}]{Deutsch_1991}%
  \BibitemOpen
  \bibfield  {author} {\bibinfo {author} {\bibfnamefont {J.~M.}\ \bibnamefont
  {Deutsch}},\ }\bibfield  {title} {\bibinfo {title} {Quantum statistical
  mechanics in a closed system},\ }\href
  {https://doi.org/10.1103/PhysRevA.43.2046} {\bibfield  {journal} {\bibinfo
  {journal} {Phys. Rev. A}\ }\textbf {\bibinfo {volume} {43}},\ \bibinfo
  {pages} {2046} (\bibinfo {year} {1991})}\BibitemShut {NoStop}%
\bibitem [{\citenamefont {Srednicki}(1994)}]{Srednicki_1994}%
  \BibitemOpen
  \bibfield  {author} {\bibinfo {author} {\bibfnamefont {M.}~\bibnamefont
  {Srednicki}},\ }\bibfield  {title} {\bibinfo {title} {Chaos and quantum
  thermalization},\ }\href {https://doi.org/10.1103/PhysRevE.50.888} {\bibfield
   {journal} {\bibinfo  {journal} {Phys. Rev. E}\ }\textbf {\bibinfo {volume}
  {50}},\ \bibinfo {pages} {888} (\bibinfo {year} {1994})}\BibitemShut
  {NoStop}%
\bibitem [{\citenamefont {Deutsch}(2018)}]{Deutsch_2018}%
  \BibitemOpen
  \bibfield  {author} {\bibinfo {author} {\bibfnamefont {J.~M.}\ \bibnamefont
  {Deutsch}},\ }\bibfield  {title} {\bibinfo {title} {Eigenstate thermalization
  hypothesis},\ }\href {https://doi.org/10.1088/1361-6633/aac9f1} {\bibfield
  {journal} {\bibinfo  {journal} {Reports on Progress in Physics}\ }\textbf
  {\bibinfo {volume} {81}},\ \bibinfo {pages} {082001} (\bibinfo {year}
  {2018})}\BibitemShut {NoStop}%
\bibitem [{\citenamefont {Rigol}\ \emph {et~al.}(2008)\citenamefont {Rigol},
  \citenamefont {Dunjko},\ and\ \citenamefont {Olshanii}}]{Rigol_2008}%
  \BibitemOpen
  \bibfield  {author} {\bibinfo {author} {\bibfnamefont {M.}~\bibnamefont
  {Rigol}}, \bibinfo {author} {\bibfnamefont {V.}~\bibnamefont {Dunjko}},\ and\
  \bibinfo {author} {\bibfnamefont {M.}~\bibnamefont {Olshanii}},\ }\bibfield
  {title} {\bibinfo {title} {Thermalization and its mechanism for generic
  isolated quantum systems},\ }\href {https://doi.org/10.1038/nature06838}
  {\bibfield  {journal} {\bibinfo  {journal} {Nature}\ }\textbf {\bibinfo
  {volume} {452}},\ \bibinfo {pages} {854} (\bibinfo {year}
  {2008})}\BibitemShut {NoStop}%
\bibitem [{\citenamefont {Rigol}\ and\ \citenamefont
  {Srednicki}(2012)}]{Rigol_2012}%
  \BibitemOpen
  \bibfield  {author} {\bibinfo {author} {\bibfnamefont {M.}~\bibnamefont
  {Rigol}}\ and\ \bibinfo {author} {\bibfnamefont {M.}~\bibnamefont
  {Srednicki}},\ }\bibfield  {title} {\bibinfo {title} {Alternatives to
  eigenstate thermalization},\ }\href
  {https://doi.org/10.1103/PhysRevLett.108.110601} {\bibfield  {journal}
  {\bibinfo  {journal} {Phys. Rev. Lett.}\ }\textbf {\bibinfo {volume} {108}},\
  \bibinfo {pages} {110601} (\bibinfo {year} {2012})}\BibitemShut {NoStop}%
\bibitem [{\citenamefont {D{\textquotesingle}Alessio}\ \emph
  {et~al.}(2016)\citenamefont {D{\textquotesingle}Alessio}, \citenamefont
  {Kafri}, \citenamefont {Polkovnikov},\ and\ \citenamefont
  {Rigol}}]{Rigol_2016}%
  \BibitemOpen
  \bibfield  {author} {\bibinfo {author} {\bibfnamefont {L.}~\bibnamefont
  {D{\textquotesingle}Alessio}}, \bibinfo {author} {\bibfnamefont
  {Y.}~\bibnamefont {Kafri}}, \bibinfo {author} {\bibfnamefont
  {A.}~\bibnamefont {Polkovnikov}},\ and\ \bibinfo {author} {\bibfnamefont
  {M.}~\bibnamefont {Rigol}},\ }\bibfield  {title} {\bibinfo {title} {From
  quantum chaos and eigenstate thermalization to statistical mechanics and
  thermodynamics},\ }\href {https://doi.org/10.1080/00018732.2016.1198134}
  {\bibfield  {journal} {\bibinfo  {journal} {Advances in Physics}\ }\textbf
  {\bibinfo {volume} {65}},\ \bibinfo {pages} {239} (\bibinfo {year}
  {2016})}\BibitemShut {NoStop}%
\bibitem [{\citenamefont {Santos}\ and\ \citenamefont
  {Rigol}(2010)}]{Santos_2010}%
  \BibitemOpen
  \bibfield  {author} {\bibinfo {author} {\bibfnamefont {L.~F.}\ \bibnamefont
  {Santos}}\ and\ \bibinfo {author} {\bibfnamefont {M.}~\bibnamefont {Rigol}},\
  }\bibfield  {title} {\bibinfo {title} {Localization and the effects of
  symmetries in the thermalization properties of one-dimensional quantum
  systems},\ }\href {https://doi.org/10.1103/PhysRevE.82.031130} {\bibfield
  {journal} {\bibinfo  {journal} {Phys. Rev. E}\ }\textbf {\bibinfo {volume}
  {82}},\ \bibinfo {pages} {031130} (\bibinfo {year} {2010})}\BibitemShut
  {NoStop}%
\bibitem [{\citenamefont {Torres-Herrera}\ and\ \citenamefont
  {Santos}(2013)}]{Santos_2013}%
  \BibitemOpen
  \bibfield  {author} {\bibinfo {author} {\bibfnamefont {E.~J.}\ \bibnamefont
  {Torres-Herrera}}\ and\ \bibinfo {author} {\bibfnamefont {L.~F.}\
  \bibnamefont {Santos}},\ }\bibfield  {title} {\bibinfo {title} {Effects of
  the interplay between initial state and hamiltonian on the thermalization of
  isolated quantum many-body systems},\ }\href
  {https://doi.org/10.1103/PhysRevE.88.042121} {\bibfield  {journal} {\bibinfo
  {journal} {Phys. Rev. E}\ }\textbf {\bibinfo {volume} {88}},\ \bibinfo
  {pages} {042121} (\bibinfo {year} {2013})}\BibitemShut {NoStop}%
\bibitem [{\citenamefont {Nandkishore}\ and\ \citenamefont
  {Huse}(2015)}]{Rahul_2015}%
  \BibitemOpen
  \bibfield  {author} {\bibinfo {author} {\bibfnamefont {R.}~\bibnamefont
  {Nandkishore}}\ and\ \bibinfo {author} {\bibfnamefont {D.~A.}\ \bibnamefont
  {Huse}},\ }\bibfield  {title} {\bibinfo {title} {Many-body localization and
  thermalization in quantum statistical mechanics},\ }\href
  {https://doi.org/10.1146/annurev-conmatphys-031214-014726} {\bibfield
  {journal} {\bibinfo  {journal} {Annual Review of Condensed Matter Physics}\
  }\textbf {\bibinfo {volume} {6}},\ \bibinfo {pages} {15} (\bibinfo {year}
  {2015})},\ \Eprint
  {https://arxiv.org/abs/https://doi.org/10.1146/annurev-conmatphys-031214-014726}
  {https://doi.org/10.1146/annurev-conmatphys-031214-014726} \BibitemShut
  {NoStop}%
\bibitem [{\citenamefont {Alet}\ and\ \citenamefont
  {Laflorencie}(2018)}]{AleLaf2018}%
  \BibitemOpen
  \bibfield  {author} {\bibinfo {author} {\bibfnamefont {F.}~\bibnamefont
  {Alet}}\ and\ \bibinfo {author} {\bibfnamefont {N.}~\bibnamefont
  {Laflorencie}},\ }\bibfield  {title} {\bibinfo {title} {Many-body
  localization: An introduction and selected topics},\ }\href
  {https://doi.org/https://doi.org/10.1016/j.crhy.2018.03.003} {\bibfield
  {journal} {\bibinfo  {journal} {Comptes Rendus Physique}\ }\textbf {\bibinfo
  {volume} {19}},\ \bibinfo {pages} {498} (\bibinfo {year} {2018})},\ \bibinfo
  {note} {quantum simulation / Simulation quantique}\BibitemShut {NoStop}%
\bibitem [{\citenamefont {Abanin}\ \emph {et~al.}(2019)\citenamefont {Abanin},
  \citenamefont {Altman}, \citenamefont {Bloch},\ and\ \citenamefont
  {Serbyn}}]{Abanin_2019}%
  \BibitemOpen
  \bibfield  {author} {\bibinfo {author} {\bibfnamefont {D.~A.}\ \bibnamefont
  {Abanin}}, \bibinfo {author} {\bibfnamefont {E.}~\bibnamefont {Altman}},
  \bibinfo {author} {\bibfnamefont {I.}~\bibnamefont {Bloch}},\ and\ \bibinfo
  {author} {\bibfnamefont {M.}~\bibnamefont {Serbyn}},\ }\bibfield  {title}
  {\bibinfo {title} {Colloquium: Many-body localization, thermalization, and
  entanglement},\ }\href {https://doi.org/10.1103/RevModPhys.91.021001}
  {\bibfield  {journal} {\bibinfo  {journal} {Rev. Mod. Phys.}\ }\textbf
  {\bibinfo {volume} {91}},\ \bibinfo {pages} {021001} (\bibinfo {year}
  {2019})}\BibitemShut {NoStop}%
\bibitem [{\citenamefont {Bernien}\ \emph {et~al.}(2017)\citenamefont
  {Bernien}, \citenamefont {Schwartz}, \citenamefont {Keesling}, \citenamefont
  {Levine}, \citenamefont {Omran}, \citenamefont {Pichler}, \citenamefont
  {Choi}, \citenamefont {Zibrov}, \citenamefont {Endres}, \citenamefont
  {Greiner}, \citenamefont {Vuleti{\'{c}}},\ and\ \citenamefont
  {Lukin}}]{BerSchKeeLev2017}%
  \BibitemOpen
  \bibfield  {author} {\bibinfo {author} {\bibfnamefont {H.}~\bibnamefont
  {Bernien}}, \bibinfo {author} {\bibfnamefont {S.}~\bibnamefont {Schwartz}},
  \bibinfo {author} {\bibfnamefont {A.}~\bibnamefont {Keesling}}, \bibinfo
  {author} {\bibfnamefont {H.}~\bibnamefont {Levine}}, \bibinfo {author}
  {\bibfnamefont {A.}~\bibnamefont {Omran}}, \bibinfo {author} {\bibfnamefont
  {H.}~\bibnamefont {Pichler}}, \bibinfo {author} {\bibfnamefont
  {S.}~\bibnamefont {Choi}}, \bibinfo {author} {\bibfnamefont {A.~S.}\
  \bibnamefont {Zibrov}}, \bibinfo {author} {\bibfnamefont {M.}~\bibnamefont
  {Endres}}, \bibinfo {author} {\bibfnamefont {M.}~\bibnamefont {Greiner}},
  \bibinfo {author} {\bibfnamefont {V.}~\bibnamefont {Vuleti{\'{c}}}},\ and\
  \bibinfo {author} {\bibfnamefont {M.~D.}\ \bibnamefont {Lukin}},\ }\bibfield
  {title} {\bibinfo {title} {Probing many-body dynamics on a 51-atom quantum
  simulator},\ }\href {https://doi.org/10.1038/nature24622} {\bibfield
  {journal} {\bibinfo  {journal} {Nature}\ }\textbf {\bibinfo {volume} {551}},\
  \bibinfo {pages} {579} (\bibinfo {year} {2017})}\BibitemShut {NoStop}%
\bibitem [{\citenamefont {Bluvstein}\ \emph {et~al.}(2021)\citenamefont
  {Bluvstein}, \citenamefont {Omran}, \citenamefont {Levine}, \citenamefont
  {Keesling}, \citenamefont {Semeghini}, \citenamefont {Ebadi}, \citenamefont
  {Wang}, \citenamefont {Michailidis}, \citenamefont {Maskara}, \citenamefont
  {Ho}, \citenamefont {Choi}, \citenamefont {Serbyn}, \citenamefont {Greiner},
  \citenamefont {Vuletić},\ and\ \citenamefont
  {Lukin}}]{BluOmrKeeSemetal2021}%
  \BibitemOpen
  \bibfield  {author} {\bibinfo {author} {\bibfnamefont {D.}~\bibnamefont
  {Bluvstein}}, \bibinfo {author} {\bibfnamefont {A.}~\bibnamefont {Omran}},
  \bibinfo {author} {\bibfnamefont {H.}~\bibnamefont {Levine}}, \bibinfo
  {author} {\bibfnamefont {A.}~\bibnamefont {Keesling}}, \bibinfo {author}
  {\bibfnamefont {G.}~\bibnamefont {Semeghini}}, \bibinfo {author}
  {\bibfnamefont {S.}~\bibnamefont {Ebadi}}, \bibinfo {author} {\bibfnamefont
  {T.~T.}\ \bibnamefont {Wang}}, \bibinfo {author} {\bibfnamefont {A.~A.}\
  \bibnamefont {Michailidis}}, \bibinfo {author} {\bibfnamefont
  {N.}~\bibnamefont {Maskara}}, \bibinfo {author} {\bibfnamefont {W.~W.}\
  \bibnamefont {Ho}}, \bibinfo {author} {\bibfnamefont {S.}~\bibnamefont
  {Choi}}, \bibinfo {author} {\bibfnamefont {M.}~\bibnamefont {Serbyn}},
  \bibinfo {author} {\bibfnamefont {M.}~\bibnamefont {Greiner}}, \bibinfo
  {author} {\bibfnamefont {V.}~\bibnamefont {Vuletić}},\ and\ \bibinfo
  {author} {\bibfnamefont {M.~D.}\ \bibnamefont {Lukin}},\ }\bibfield  {title}
  {\bibinfo {title} {Controlling quantum many-body dynamics in driven rydberg
  atom arrays},\ }\href {https://doi.org/10.1126/science.abg2530} {\bibfield
  {journal} {\bibinfo  {journal} {Science}\ }\textbf {\bibinfo {volume}
  {371}},\ \bibinfo {pages} {1355} (\bibinfo {year} {2021})},\ \Eprint
  {https://arxiv.org/abs/https://www.science.org/doi/pdf/10.1126/science.abg2530}
  {https://www.science.org/doi/pdf/10.1126/science.abg2530} \BibitemShut
  {NoStop}%
\bibitem [{\citenamefont {Calabrese}\ and\ \citenamefont
  {Cardy}(2005{\natexlab{a}})}]{Calabrese_2005}%
  \BibitemOpen
  \bibfield  {author} {\bibinfo {author} {\bibfnamefont {P.}~\bibnamefont
  {Calabrese}}\ and\ \bibinfo {author} {\bibfnamefont {J.}~\bibnamefont
  {Cardy}},\ }\bibfield  {title} {\bibinfo {title} {Evolution of entanglement
  entropy in one-dimensional systems},\ }\href
  {https://doi.org/10.1088/1742-5468/2005/04/p04010} {\bibfield  {journal}
  {\bibinfo  {journal} {Journal of Statistical Mechanics: Theory and
  Experiment}\ }\textbf {\bibinfo {volume} {2005}},\ \bibinfo {pages} {P04010}
  (\bibinfo {year} {2005}{\natexlab{a}})}\BibitemShut {NoStop}%
\bibitem [{\citenamefont {Calabrese}\ and\ \citenamefont
  {Cardy}(2016)}]{Calabrese_2016}%
  \BibitemOpen
  \bibfield  {author} {\bibinfo {author} {\bibfnamefont {P.}~\bibnamefont
  {Calabrese}}\ and\ \bibinfo {author} {\bibfnamefont {J.}~\bibnamefont
  {Cardy}},\ }\bibfield  {title} {\bibinfo {title} {Quantum quenches in 1+1
  dimensional conformal field theories},\ }\href
  {https://doi.org/10.1088/1742-5468/2016/06/064003} {\bibfield  {journal}
  {\bibinfo  {journal} {Journal of Statistical Mechanics: Theory and
  Experiment}\ }\textbf {\bibinfo {volume} {2016}},\ \bibinfo {pages} {064003}
  (\bibinfo {year} {2016})}\BibitemShut {NoStop}%
\bibitem [{\citenamefont {Eisler}\ and\ \citenamefont
  {Peschel}(2008)}]{Eisler_2008}%
  \BibitemOpen
  \bibfield  {author} {\bibinfo {author} {\bibfnamefont {V.}~\bibnamefont
  {Eisler}}\ and\ \bibinfo {author} {\bibfnamefont {I.}~\bibnamefont
  {Peschel}},\ }\bibfield  {title} {\bibinfo {title} {Entanglement in a
  periodic quench},\ }\href {https://doi.org/10.1002/andp.20085200605}
  {\bibfield  {journal} {\bibinfo  {journal} {Annalen der Physik}\ }\textbf
  {\bibinfo {volume} {520}},\ \bibinfo {pages} {410} (\bibinfo {year}
  {2008})}\BibitemShut {NoStop}%
\bibitem [{\citenamefont {Alba}(2018)}]{Alba_2018}%
  \BibitemOpen
  \bibfield  {author} {\bibinfo {author} {\bibfnamefont {V.}~\bibnamefont
  {Alba}},\ }\bibfield  {title} {\bibinfo {title} {Entanglement and quantum
  transport in integrable systems},\ }\href
  {https://doi.org/10.1103/PhysRevB.97.245135} {\bibfield  {journal} {\bibinfo
  {journal} {Phys. Rev. B}\ }\textbf {\bibinfo {volume} {97}},\ \bibinfo
  {pages} {245135} (\bibinfo {year} {2018})}\BibitemShut {NoStop}%
\bibitem [{\citenamefont {Chiara}\ \emph {et~al.}(2006)\citenamefont {Chiara},
  \citenamefont {Montangero}, \citenamefont {Calabrese},\ and\ \citenamefont
  {Fazio}}]{Chiara_2006}%
  \BibitemOpen
  \bibfield  {author} {\bibinfo {author} {\bibfnamefont {G.~D.}\ \bibnamefont
  {Chiara}}, \bibinfo {author} {\bibfnamefont {S.}~\bibnamefont {Montangero}},
  \bibinfo {author} {\bibfnamefont {P.}~\bibnamefont {Calabrese}},\ and\
  \bibinfo {author} {\bibfnamefont {R.}~\bibnamefont {Fazio}},\ }\bibfield
  {title} {\bibinfo {title} {Entanglement entropy dynamics of heisenberg
  chains},\ }\href {https://doi.org/10.1088/1742-5468/2006/03/p03001}
  {\bibfield  {journal} {\bibinfo  {journal} {Journal of Statistical Mechanics:
  Theory and Experiment}\ }\textbf {\bibinfo {volume} {2006}},\ \bibinfo
  {pages} {P03001} (\bibinfo {year} {2006})}\BibitemShut {NoStop}%
\bibitem [{\citenamefont {Calabrese}(2020)}]{Calabrese_2020}%
  \BibitemOpen
  \bibfield  {author} {\bibinfo {author} {\bibfnamefont {P.}~\bibnamefont
  {Calabrese}},\ }\bibfield  {title} {\bibinfo {title} {Entanglement spreading
  in non-equilibrium integrable systems},\ }\bibfield  {journal} {\bibinfo
  {journal} {{SciPost} Physics Lecture Notes}\ }\href
  {https://doi.org/10.21468/scipostphyslectnotes.20}
  {10.21468/scipostphyslectnotes.20} (\bibinfo {year} {2020})\BibitemShut
  {NoStop}%
\bibitem [{\citenamefont {Fagotti}\ and\ \citenamefont
  {Calabrese}(2008)}]{Fagotti_2008}%
  \BibitemOpen
  \bibfield  {author} {\bibinfo {author} {\bibfnamefont {M.}~\bibnamefont
  {Fagotti}}\ and\ \bibinfo {author} {\bibfnamefont {P.}~\bibnamefont
  {Calabrese}},\ }\bibfield  {title} {\bibinfo {title} {Evolution of
  entanglement entropy following a quantum quench: Analytic results for the
  $xy$ chain in a transverse magnetic field},\ }\href
  {https://doi.org/10.1103/PhysRevA.78.010306} {\bibfield  {journal} {\bibinfo
  {journal} {Phys. Rev. A}\ }\textbf {\bibinfo {volume} {78}},\ \bibinfo
  {pages} {010306(R)} (\bibinfo {year} {2008})}\BibitemShut {NoStop}%
\bibitem [{\citenamefont {Nezhadhaghighi}\ and\ \citenamefont
  {Rajabpour}(2014)}]{Nezgharaja_2014}%
  \BibitemOpen
  \bibfield  {author} {\bibinfo {author} {\bibfnamefont {M.~G.}\ \bibnamefont
  {Nezhadhaghighi}}\ and\ \bibinfo {author} {\bibfnamefont {M.~A.}\
  \bibnamefont {Rajabpour}},\ }\bibfield  {title} {\bibinfo {title}
  {Entanglement dynamics in short- and long-range harmonic oscillators},\
  }\href {https://doi.org/10.1103/PhysRevB.90.205438} {\bibfield  {journal}
  {\bibinfo  {journal} {Phys. Rev. B}\ }\textbf {\bibinfo {volume} {90}},\
  \bibinfo {pages} {205438} (\bibinfo {year} {2014})}\BibitemShut {NoStop}%
\bibitem [{\citenamefont {Coser}\ \emph {et~al.}(2014)\citenamefont {Coser},
  \citenamefont {Tonni},\ and\ \citenamefont {Calabrese}}]{Coser_2014}%
  \BibitemOpen
  \bibfield  {author} {\bibinfo {author} {\bibfnamefont {A.}~\bibnamefont
  {Coser}}, \bibinfo {author} {\bibfnamefont {E.}~\bibnamefont {Tonni}},\ and\
  \bibinfo {author} {\bibfnamefont {P.}~\bibnamefont {Calabrese}},\ }\bibfield
  {title} {\bibinfo {title} {Entanglement negativity after a global quantum
  quench},\ }\href {https://doi.org/10.1088/1742-5468/2014/12/p12017}
  {\bibfield  {journal} {\bibinfo  {journal} {Journal of Statistical Mechanics:
  Theory and Experiment}\ }\textbf {\bibinfo {volume} {2014}},\ \bibinfo
  {pages} {P12017} (\bibinfo {year} {2014})}\BibitemShut {NoStop}%
\bibitem [{\citenamefont {Buyskikh}\ \emph {et~al.}(2016)\citenamefont
  {Buyskikh}, \citenamefont {Fagotti}, \citenamefont {Schachenmayer},
  \citenamefont {Essler},\ and\ \citenamefont
  {Daley}}]{Buyantmauessfaband_2016}%
  \BibitemOpen
  \bibfield  {author} {\bibinfo {author} {\bibfnamefont {A.~S.}\ \bibnamefont
  {Buyskikh}}, \bibinfo {author} {\bibfnamefont {M.}~\bibnamefont {Fagotti}},
  \bibinfo {author} {\bibfnamefont {J.}~\bibnamefont {Schachenmayer}}, \bibinfo
  {author} {\bibfnamefont {F.}~\bibnamefont {Essler}},\ and\ \bibinfo {author}
  {\bibfnamefont {A.~J.}\ \bibnamefont {Daley}},\ }\bibfield  {title} {\bibinfo
  {title} {Entanglement growth and correlation spreading with variable-range
  interactions in spin and fermionic tunneling models},\ }\href
  {https://doi.org/10.1103/PhysRevA.93.053620} {\bibfield  {journal} {\bibinfo
  {journal} {Phys. Rev. A}\ }\textbf {\bibinfo {volume} {93}},\ \bibinfo
  {pages} {053620} (\bibinfo {year} {2016})}\BibitemShut {NoStop}%
\bibitem [{\citenamefont {Kudler-Flam}\ \emph {et~al.}(2021)\citenamefont
  {Kudler-Flam}, \citenamefont {Kusuki},\ and\ \citenamefont
  {Ryu}}]{Kudler_Flam_2021}%
  \BibitemOpen
  \bibfield  {author} {\bibinfo {author} {\bibfnamefont {J.}~\bibnamefont
  {Kudler-Flam}}, \bibinfo {author} {\bibfnamefont {Y.}~\bibnamefont
  {Kusuki}},\ and\ \bibinfo {author} {\bibfnamefont {S.}~\bibnamefont {Ryu}},\
  }\bibfield  {title} {\bibinfo {title} {The quasi-particle picture and its
  breakdown after local quenches: mutual information, negativity, and reflected
  entropy},\ }\bibfield  {journal} {\bibinfo  {journal} {Journal of High Energy
  Physics}\ }\textbf {\bibinfo {volume} {2021}},\ \href
  {https://doi.org/10.1007/jhep03(2021)146} {10.1007/jhep03(2021)146} (\bibinfo
  {year} {2021})\BibitemShut {NoStop}%
\bibitem [{\citenamefont {Alba}\ and\ \citenamefont
  {Calabrese}(2017)}]{Alba_2017}%
  \BibitemOpen
  \bibfield  {author} {\bibinfo {author} {\bibfnamefont {V.}~\bibnamefont
  {Alba}}\ and\ \bibinfo {author} {\bibfnamefont {P.}~\bibnamefont
  {Calabrese}},\ }\bibfield  {title} {\bibinfo {title} {Entanglement and
  thermodynamics after a quantum quench in integrable systems},\ }\href
  {https://doi.org/10.1073/pnas.1703516114} {\bibfield  {journal} {\bibinfo
  {journal} {Proceedings of the National Academy of Sciences}\ }\textbf
  {\bibinfo {volume} {114}},\ \bibinfo {pages} {7947} (\bibinfo {year}
  {2017})}\BibitemShut {NoStop}%
\bibitem [{\citenamefont {Alba}\ \emph {et~al.}(2021)\citenamefont {Alba},
  \citenamefont {Bertini}, \citenamefont {Fagotti}, \citenamefont {Piroli},\
  and\ \citenamefont {Ruggiero}}]{Alba_2021}%
  \BibitemOpen
  \bibfield  {author} {\bibinfo {author} {\bibfnamefont {V.}~\bibnamefont
  {Alba}}, \bibinfo {author} {\bibfnamefont {B.}~\bibnamefont {Bertini}},
  \bibinfo {author} {\bibfnamefont {M.}~\bibnamefont {Fagotti}}, \bibinfo
  {author} {\bibfnamefont {L.}~\bibnamefont {Piroli}},\ and\ \bibinfo {author}
  {\bibfnamefont {P.}~\bibnamefont {Ruggiero}},\ }\bibfield  {title} {\bibinfo
  {title} {Generalized-hydrodynamic approach to inhomogeneous quenches:
  correlations, entanglement and quantum effects},\ }\href
  {https://doi.org/10.1088/1742-5468/ac257d} {\bibfield  {journal} {\bibinfo
  {journal} {Journal of Statistical Mechanics: Theory and Experiment}\ }\textbf
  {\bibinfo {volume} {2021}},\ \bibinfo {pages} {114004} (\bibinfo {year}
  {2021})}\BibitemShut {NoStop}%
\bibitem [{\citenamefont {Lakshminarayan}\ \emph {et~al.}(2016)\citenamefont
  {Lakshminarayan}, \citenamefont {Srivastava}, \citenamefont {Ketzmerick},
  \citenamefont {B\"acker},\ and\ \citenamefont {Tomsovic}}]{Arul_2016}%
  \BibitemOpen
  \bibfield  {author} {\bibinfo {author} {\bibfnamefont {A.}~\bibnamefont
  {Lakshminarayan}}, \bibinfo {author} {\bibfnamefont {S.~C.~L.}\ \bibnamefont
  {Srivastava}}, \bibinfo {author} {\bibfnamefont {R.}~\bibnamefont
  {Ketzmerick}}, \bibinfo {author} {\bibfnamefont {A.}~\bibnamefont
  {B\"acker}},\ and\ \bibinfo {author} {\bibfnamefont {S.}~\bibnamefont
  {Tomsovic}},\ }\bibfield  {title} {\bibinfo {title} {Entanglement and
  localization transitions in eigenstates of interacting chaotic systems},\
  }\href {https://doi.org/10.1103/PhysRevE.94.010205} {\bibfield  {journal}
  {\bibinfo  {journal} {Phys. Rev. E}\ }\textbf {\bibinfo {volume} {94}},\
  \bibinfo {pages} {010205(R)} (\bibinfo {year} {2016})}\BibitemShut {NoStop}%
\bibitem [{\citenamefont {Lakshminarayan}(2001)}]{Arul_2001}%
  \BibitemOpen
  \bibfield  {author} {\bibinfo {author} {\bibfnamefont {A.}~\bibnamefont
  {Lakshminarayan}},\ }\bibfield  {title} {\bibinfo {title} {Entangling power
  of quantized chaotic systems},\ }\href
  {https://doi.org/10.1103/PhysRevE.64.036207} {\bibfield  {journal} {\bibinfo
  {journal} {Phys. Rev. E}\ }\textbf {\bibinfo {volume} {64}},\ \bibinfo
  {pages} {036207} (\bibinfo {year} {2001})}\BibitemShut {NoStop}%
\bibitem [{\citenamefont {Bandyopadhyay}\ and\ \citenamefont
  {Lakshminarayan}(2002)}]{Arul_2002}%
  \BibitemOpen
  \bibfield  {author} {\bibinfo {author} {\bibfnamefont {J.~N.}\ \bibnamefont
  {Bandyopadhyay}}\ and\ \bibinfo {author} {\bibfnamefont {A.}~\bibnamefont
  {Lakshminarayan}},\ }\bibfield  {title} {\bibinfo {title} {Testing
  statistical bounds on entanglement using quantum chaos},\ }\href
  {https://doi.org/10.1103/PhysRevLett.89.060402} {\bibfield  {journal}
  {\bibinfo  {journal} {Phys. Rev. Lett.}\ }\textbf {\bibinfo {volume} {89}},\
  \bibinfo {pages} {060402} (\bibinfo {year} {2002})}\BibitemShut {NoStop}%
\bibitem [{\citenamefont {Fujisaki}\ \emph {et~al.}(2003)\citenamefont
  {Fujisaki}, \citenamefont {Tanaka},\ and\ \citenamefont
  {Miyadera}}]{Fujisaki_2003}%
  \BibitemOpen
  \bibfield  {author} {\bibinfo {author} {\bibfnamefont {H.}~\bibnamefont
  {Fujisaki}}, \bibinfo {author} {\bibfnamefont {A.}~\bibnamefont {Tanaka}},\
  and\ \bibinfo {author} {\bibfnamefont {T.}~\bibnamefont {Miyadera}},\
  }\bibfield  {title} {\bibinfo {title} {Dynamical aspects of quantum
  entanglement for coupled mapping systems},\ }\href
  {https://doi.org/10.1143/jpsjs.72sc.111} {\bibfield  {journal} {\bibinfo
  {journal} {Journal of the Physical Society of Japan}\ }\textbf {\bibinfo
  {volume} {72}},\ \bibinfo {pages} {111} (\bibinfo {year} {2003})}\BibitemShut
  {NoStop}%
\bibitem [{\citenamefont {Srivastava}\ \emph {et~al.}(2016)\citenamefont
  {Srivastava}, \citenamefont {Tomsovic}, \citenamefont {Lakshminarayan},
  \citenamefont {Ketzmerick},\ and\ \citenamefont {B\"acker}}]{Shashi_2016}%
  \BibitemOpen
  \bibfield  {author} {\bibinfo {author} {\bibfnamefont {S.~C.~L.}\
  \bibnamefont {Srivastava}}, \bibinfo {author} {\bibfnamefont
  {S.}~\bibnamefont {Tomsovic}}, \bibinfo {author} {\bibfnamefont
  {A.}~\bibnamefont {Lakshminarayan}}, \bibinfo {author} {\bibfnamefont
  {R.}~\bibnamefont {Ketzmerick}},\ and\ \bibinfo {author} {\bibfnamefont
  {A.}~\bibnamefont {B\"acker}},\ }\bibfield  {title} {\bibinfo {title}
  {Universal scaling of spectral fluctuation transitions for interacting
  chaotic systems},\ }\href {https://doi.org/10.1103/PhysRevLett.116.054101}
  {\bibfield  {journal} {\bibinfo  {journal} {Phys. Rev. Lett.}\ }\textbf
  {\bibinfo {volume} {116}},\ \bibinfo {pages} {054101} (\bibinfo {year}
  {2016})}\BibitemShut {NoStop}%
\bibitem [{\citenamefont {Tomsovic}\ \emph {et~al.}(2018)\citenamefont
  {Tomsovic}, \citenamefont {Lakshminarayan}, \citenamefont {Srivastava},\ and\
  \citenamefont {B\"acker}}]{Steven_2018}%
  \BibitemOpen
  \bibfield  {author} {\bibinfo {author} {\bibfnamefont {S.}~\bibnamefont
  {Tomsovic}}, \bibinfo {author} {\bibfnamefont {A.}~\bibnamefont
  {Lakshminarayan}}, \bibinfo {author} {\bibfnamefont {S.~C.~L.}\ \bibnamefont
  {Srivastava}},\ and\ \bibinfo {author} {\bibfnamefont {A.}~\bibnamefont
  {B\"acker}},\ }\bibfield  {title} {\bibinfo {title} {Eigenstate entanglement
  between quantum chaotic subsystems: Universal transitions and power laws in
  the entanglement spectrum},\ }\href
  {https://doi.org/10.1103/PhysRevE.98.032209} {\bibfield  {journal} {\bibinfo
  {journal} {Phys. Rev. E}\ }\textbf {\bibinfo {volume} {98}},\ \bibinfo
  {pages} {032209} (\bibinfo {year} {2018})}\BibitemShut {NoStop}%
\bibitem [{\citenamefont {Pulikkottil}\ \emph {et~al.}(2020)\citenamefont
  {Pulikkottil}, \citenamefont {Lakshminarayan}, \citenamefont {Srivastava},
  \citenamefont {B\"acker},\ and\ \citenamefont {Tomsovic}}]{Pulikkottil_2020}%
  \BibitemOpen
  \bibfield  {author} {\bibinfo {author} {\bibfnamefont {J.~J.}\ \bibnamefont
  {Pulikkottil}}, \bibinfo {author} {\bibfnamefont {A.}~\bibnamefont
  {Lakshminarayan}}, \bibinfo {author} {\bibfnamefont {S.~C.~L.}\ \bibnamefont
  {Srivastava}}, \bibinfo {author} {\bibfnamefont {A.}~\bibnamefont
  {B\"acker}},\ and\ \bibinfo {author} {\bibfnamefont {S.}~\bibnamefont
  {Tomsovic}},\ }\bibfield  {title} {\bibinfo {title} {Entanglement production
  by interaction quenches of quantum chaotic subsystems},\ }\href
  {https://doi.org/10.1103/PhysRevE.101.032212} {\bibfield  {journal} {\bibinfo
   {journal} {Phys. Rev. E}\ }\textbf {\bibinfo {volume} {101}},\ \bibinfo
  {pages} {032212} (\bibinfo {year} {2020})}\BibitemShut {NoStop}%
\bibitem [{\citenamefont {Pulikkottil}\ \emph {et~al.}(2023)\citenamefont
  {Pulikkottil}, \citenamefont {Lakshminarayan}, \citenamefont {Srivastava},
  \citenamefont {Kieler}, \citenamefont {B\"acker},\ and\ \citenamefont
  {Tomsovic}}]{Arul_2022}%
  \BibitemOpen
  \bibfield  {author} {\bibinfo {author} {\bibfnamefont {J.~J.}\ \bibnamefont
  {Pulikkottil}}, \bibinfo {author} {\bibfnamefont {A.}~\bibnamefont
  {Lakshminarayan}}, \bibinfo {author} {\bibfnamefont {S.~C.~L.}\ \bibnamefont
  {Srivastava}}, \bibinfo {author} {\bibfnamefont {M.~F.~I.}\ \bibnamefont
  {Kieler}}, \bibinfo {author} {\bibfnamefont {A.}~\bibnamefont {B\"acker}},\
  and\ \bibinfo {author} {\bibfnamefont {S.}~\bibnamefont {Tomsovic}},\
  }\bibfield  {title} {\bibinfo {title} {Quantum coherence controls the nature
  of equilibration and thermalization in coupled chaotic systems},\ }\href
  {https://doi.org/10.1103/PhysRevE.107.024124} {\bibfield  {journal} {\bibinfo
   {journal} {Phys. Rev. E}\ }\textbf {\bibinfo {volume} {107}},\ \bibinfo
  {pages} {024124} (\bibinfo {year} {2023})}\BibitemShut {NoStop}%
\bibitem [{\citenamefont {Lubkin}(1978)}]{Lub1978}%
  \BibitemOpen
  \bibfield  {author} {\bibinfo {author} {\bibfnamefont {E.}~\bibnamefont
  {Lubkin}},\ }\bibfield  {title} {\bibinfo {title} {Entropy of an n‐system
  from its correlation with a k‐reservoir},\ }\href
  {https://doi.org/10.1063/1.523763} {\bibfield  {journal} {\bibinfo  {journal}
  {Journal of Mathematical Physics}\ }\textbf {\bibinfo {volume} {19}},\
  \bibinfo {pages} {1028} (\bibinfo {year} {1978})}\BibitemShut {NoStop}%
\bibitem [{\citenamefont {Franchi}\ \emph {et~al.}(2022)\citenamefont
  {Franchi}, \citenamefont {Rossini},\ and\ \citenamefont
  {Vicari}}]{Franchi_2022}%
  \BibitemOpen
  \bibfield  {author} {\bibinfo {author} {\bibfnamefont {A.}~\bibnamefont
  {Franchi}}, \bibinfo {author} {\bibfnamefont {D.}~\bibnamefont {Rossini}},\
  and\ \bibinfo {author} {\bibfnamefont {E.}~\bibnamefont {Vicari}},\
  }\bibfield  {title} {\bibinfo {title} {Quantum many-body spin rings coupled
  to ancillary spins: The sunburst quantum ising model},\ }\href
  {https://doi.org/10.1103/PhysRevE.105.054111} {\bibfield  {journal} {\bibinfo
   {journal} {Phys. Rev. E}\ }\textbf {\bibinfo {volume} {105}},\ \bibinfo
  {pages} {054111} (\bibinfo {year} {2022})}\BibitemShut {NoStop}%
\bibitem [{\citenamefont {Pal}\ and\ \citenamefont {Huse}(2010)}]{PalHus2010}%
  \BibitemOpen
  \bibfield  {author} {\bibinfo {author} {\bibfnamefont {A.}~\bibnamefont
  {Pal}}\ and\ \bibinfo {author} {\bibfnamefont {D.~A.}\ \bibnamefont {Huse}},\
  }\bibfield  {title} {\bibinfo {title} {Many-body localization phase
  transition},\ }\href {https://doi.org/10.1103/PhysRevB.82.174411} {\bibfield
  {journal} {\bibinfo  {journal} {Phys. Rev. B}\ }\textbf {\bibinfo {volume}
  {82}},\ \bibinfo {pages} {174411} (\bibinfo {year} {2010})}\BibitemShut
  {NoStop}%
\bibitem [{\citenamefont {Luitz}\ \emph {et~al.}(2015)\citenamefont {Luitz},
  \citenamefont {Laflorencie},\ and\ \citenamefont {Alet}}]{LuiLafFab2015}%
  \BibitemOpen
  \bibfield  {author} {\bibinfo {author} {\bibfnamefont {D.~J.}\ \bibnamefont
  {Luitz}}, \bibinfo {author} {\bibfnamefont {N.}~\bibnamefont {Laflorencie}},\
  and\ \bibinfo {author} {\bibfnamefont {F.}~\bibnamefont {Alet}},\ }\bibfield
  {title} {\bibinfo {title} {Many-body localization edge in the random-field
  heisenberg chain},\ }\href {https://doi.org/10.1103/PhysRevB.91.081103}
  {\bibfield  {journal} {\bibinfo  {journal} {Phys. Rev. B}\ }\textbf {\bibinfo
  {volume} {91}},\ \bibinfo {pages} {081103(R)} (\bibinfo {year}
  {2015})}\BibitemShut {NoStop}%
\bibitem [{\citenamefont {Serbyn}\ \emph {et~al.}(2015)\citenamefont {Serbyn},
  \citenamefont {Papi\ifmmode~\acute{c}\else \'{c}\fi{}},\ and\ \citenamefont
  {Abanin}}]{SerPapAba2015}%
  \BibitemOpen
  \bibfield  {author} {\bibinfo {author} {\bibfnamefont {M.}~\bibnamefont
  {Serbyn}}, \bibinfo {author} {\bibfnamefont {Z.}~\bibnamefont
  {Papi\ifmmode~\acute{c}\else \'{c}\fi{}}},\ and\ \bibinfo {author}
  {\bibfnamefont {D.~A.}\ \bibnamefont {Abanin}},\ }\bibfield  {title}
  {\bibinfo {title} {Criterion for many-body localization-delocalization phase
  transition},\ }\href {https://doi.org/10.1103/PhysRevX.5.041047} {\bibfield
  {journal} {\bibinfo  {journal} {Phys. Rev. X}\ }\textbf {\bibinfo {volume}
  {5}},\ \bibinfo {pages} {041047} (\bibinfo {year} {2015})}\BibitemShut
  {NoStop}%
\bibitem [{\citenamefont {Bera}\ and\ \citenamefont
  {Lakshminarayan}(2016)}]{BerLak2016}%
  \BibitemOpen
  \bibfield  {author} {\bibinfo {author} {\bibfnamefont {S.}~\bibnamefont
  {Bera}}\ and\ \bibinfo {author} {\bibfnamefont {A.}~\bibnamefont
  {Lakshminarayan}},\ }\bibfield  {title} {\bibinfo {title} {Local entanglement
  structure across a many-body localization transition},\ }\href
  {https://doi.org/10.1103/PhysRevB.93.134204} {\bibfield  {journal} {\bibinfo
  {journal} {Phys. Rev. B}\ }\textbf {\bibinfo {volume} {93}},\ \bibinfo
  {pages} {134204} (\bibinfo {year} {2016})}\BibitemShut {NoStop}%
\bibitem [{\citenamefont {Berry}\ and\ \citenamefont {Tabor}(1977)}]{Berry77b}%
  \BibitemOpen
  \bibfield  {author} {\bibinfo {author} {\bibfnamefont {M.~V.}\ \bibnamefont
  {Berry}}\ and\ \bibinfo {author} {\bibfnamefont {M.}~\bibnamefont {Tabor}},\
  }\bibfield  {title} {\bibinfo {title} {Level clustering in the regular
  spectrum},\ }\href@noop {} {\bibfield  {journal} {\bibinfo  {journal}
  {Proc.~R.~Soc.~A}\ }\textbf {\bibinfo {volume} {356}},\ \bibinfo {pages}
  {375} (\bibinfo {year} {1977})}\BibitemShut {NoStop}%
\bibitem [{\citenamefont {Bohigas}\ \emph {et~al.}(1984)\citenamefont
  {Bohigas}, \citenamefont {Giannoni},\ and\ \citenamefont
  {Schmit}}]{Bohigas84}%
  \BibitemOpen
  \bibfield  {author} {\bibinfo {author} {\bibfnamefont {O.}~\bibnamefont
  {Bohigas}}, \bibinfo {author} {\bibfnamefont {M.-J.}\ \bibnamefont
  {Giannoni}},\ and\ \bibinfo {author} {\bibfnamefont {C.}~\bibnamefont
  {Schmit}},\ }\bibfield  {title} {\bibinfo {title} {Characterization of
  chaotic quantum spectra and universality of level fluctuation laws},\
  }\href@noop {} {\bibfield  {journal} {\bibinfo  {journal} {Physical Review
  Letters}\ }\textbf {\bibinfo {volume} {52}},\ \bibinfo {pages} {1} (\bibinfo
  {year} {1984})}\BibitemShut {NoStop}%
\bibitem [{\citenamefont {Oganesyan}\ and\ \citenamefont
  {Huse}(2007)}]{OgaHus2007}%
  \BibitemOpen
  \bibfield  {author} {\bibinfo {author} {\bibfnamefont {V.}~\bibnamefont
  {Oganesyan}}\ and\ \bibinfo {author} {\bibfnamefont {D.~A.}\ \bibnamefont
  {Huse}},\ }\bibfield  {title} {\bibinfo {title} {Localization of interacting
  fermions at high temperature},\ }\href
  {https://doi.org/10.1103/PhysRevB.75.155111} {\bibfield  {journal} {\bibinfo
  {journal} {Phys. Rev. B}\ }\textbf {\bibinfo {volume} {75}},\ \bibinfo
  {pages} {155111} (\bibinfo {year} {2007})}\BibitemShut {NoStop}%
\bibitem [{\citenamefont {Atas}\ \emph {et~al.}(2013)\citenamefont {Atas},
  \citenamefont {Bogomolny}, \citenamefont {Giraud},\ and\ \citenamefont
  {Roux}}]{AtaBogGirRou2013}%
  \BibitemOpen
  \bibfield  {author} {\bibinfo {author} {\bibfnamefont {Y.~Y.}\ \bibnamefont
  {Atas}}, \bibinfo {author} {\bibfnamefont {E.}~\bibnamefont {Bogomolny}},
  \bibinfo {author} {\bibfnamefont {O.}~\bibnamefont {Giraud}},\ and\ \bibinfo
  {author} {\bibfnamefont {G.}~\bibnamefont {Roux}},\ }\bibfield  {title}
  {\bibinfo {title} {Distribution of the ratio of consecutive level spacings in
  random matrix ensembles},\ }\href
  {https://doi.org/10.1103/PhysRevLett.110.084101} {\bibfield  {journal}
  {\bibinfo  {journal} {Phys. Rev. Lett.}\ }\textbf {\bibinfo {volume} {110}},\
  \bibinfo {pages} {084101} (\bibinfo {year} {2013})}\BibitemShut {NoStop}%
\bibitem [{\citenamefont {Shnirelman}(1975)}]{Shn1975}%
  \BibitemOpen
  \bibfield  {author} {\bibinfo {author} {\bibfnamefont {A.~I.}\ \bibnamefont
  {Shnirelman}},\ }\bibfield  {title} {\bibinfo {title} {On asymptotic
  multiplicity of spectrum of laplace operator},\ }\href@noop {} {\bibfield
  {journal} {\bibinfo  {journal} {Usp. Mat. Nauk}\ }\textbf {\bibinfo {volume}
  {30}},\ \bibinfo {pages} {265} (\bibinfo {year} {1975})}\BibitemShut
  {NoStop}%
\bibitem [{\citenamefont {Shnirelman}(1993)}]{Shn1993}%
  \BibitemOpen
  \bibfield  {author} {\bibinfo {author} {\bibfnamefont {A.~I.}\ \bibnamefont
  {Shnirelman}},\ }\bibinfo {title} {Addendum on the asymptotic properties of
  eigenfunctions in the regions of chaotic motion},\ in\ \href
  {https://doi.org/10.1007/978-3-642-76247-5_10} {\emph {\bibinfo {booktitle}
  {KAM Theory and Semiclassical Approximations to Eigenfunctions}}}\ (\bibinfo
  {publisher} {Springer Berlin Heidelberg},\ \bibinfo {address} {Berlin,
  Heidelberg},\ \bibinfo {year} {1993})\ pp.\ \bibinfo {pages}
  {313--337}\BibitemShut {NoStop}%
\bibitem [{\citenamefont {Chirikov}\ and\ \citenamefont
  {Shepelyansky}(1995)}]{ChiShe1995}%
  \BibitemOpen
  \bibfield  {author} {\bibinfo {author} {\bibfnamefont {B.~V.}\ \bibnamefont
  {Chirikov}}\ and\ \bibinfo {author} {\bibfnamefont {D.~L.}\ \bibnamefont
  {Shepelyansky}},\ }\bibfield  {title} {\bibinfo {title} {Shnirelman peak in
  level spacing statistics},\ }\href
  {https://doi.org/10.1103/PhysRevLett.74.518} {\bibfield  {journal} {\bibinfo
  {journal} {Phys. Rev. Lett.}\ }\textbf {\bibinfo {volume} {74}},\ \bibinfo
  {pages} {518} (\bibinfo {year} {1995})}\BibitemShut {NoStop}%
\bibitem [{\citenamefont {Bandyopadhyay}\ and\ \citenamefont
  {Lakshminarayan}(2005)}]{Bandyopadhyay_2005}%
  \BibitemOpen
  \bibfield  {author} {\bibinfo {author} {\bibfnamefont {J.~N.}\ \bibnamefont
  {Bandyopadhyay}}\ and\ \bibinfo {author} {\bibfnamefont {A.}~\bibnamefont
  {Lakshminarayan}},\ }\bibfield  {title} {\bibinfo {title} {Entanglement
  production in quantized chaotic systems},\ }\href
  {https://doi.org/10.1007/bf02706205} {\bibfield  {journal} {\bibinfo
  {journal} {Pramana}\ }\textbf {\bibinfo {volume} {64}},\ \bibinfo {pages}
  {577} (\bibinfo {year} {2005})}\BibitemShut {NoStop}%
\bibitem [{\citenamefont {Schr{\"o}dinger}(1935)}]{Sch1935}%
  \BibitemOpen
  \bibfield  {author} {\bibinfo {author} {\bibfnamefont {E.}~\bibnamefont
  {Schr{\"o}dinger}},\ }\bibfield  {title} {\bibinfo {title} {Die
  gegenw{\"a}rtige situation in der quantenmechanik},\ }\href
  {https://doi.org/10.1007/BF01491914} {\bibfield  {journal} {\bibinfo
  {journal} {Naturwissenschaften}\ }\textbf {\bibinfo {volume} {23}},\ \bibinfo
  {pages} {823} (\bibinfo {year} {1935})}\BibitemShut {NoStop}%
\bibitem [{\citenamefont {Greenberger}\ \emph {et~al.}(2007)\citenamefont
  {Greenberger}, \citenamefont {Horne},\ and\ \citenamefont
  {Zeilinger}}]{GHZ2007}%
  \BibitemOpen
  \bibfield  {author} {\bibinfo {author} {\bibfnamefont {D.~M.}\ \bibnamefont
  {Greenberger}}, \bibinfo {author} {\bibfnamefont {M.~A.}\ \bibnamefont
  {Horne}},\ and\ \bibinfo {author} {\bibfnamefont {A.}~\bibnamefont
  {Zeilinger}},\ }\href@noop {} {\bibinfo {title} {Going beyond bell's
  theorem}} (\bibinfo {year} {2007}),\ \Eprint
  {https://arxiv.org/abs/0712.0921} {arXiv:0712.0921 [quant-ph]} \BibitemShut
  {NoStop}%
\bibitem [{\citenamefont {Bouwmeester}\ \emph {et~al.}(1999)\citenamefont
  {Bouwmeester}, \citenamefont {Pan}, \citenamefont {Daniell}, \citenamefont
  {Weinfurter},\ and\ \citenamefont {Zeilinger}}]{Zeilingeretal1999}%
  \BibitemOpen
  \bibfield  {author} {\bibinfo {author} {\bibfnamefont {D.}~\bibnamefont
  {Bouwmeester}}, \bibinfo {author} {\bibfnamefont {J.-W.}\ \bibnamefont
  {Pan}}, \bibinfo {author} {\bibfnamefont {M.}~\bibnamefont {Daniell}},
  \bibinfo {author} {\bibfnamefont {H.}~\bibnamefont {Weinfurter}},\ and\
  \bibinfo {author} {\bibfnamefont {A.}~\bibnamefont {Zeilinger}},\ }\bibfield
  {title} {\bibinfo {title} {Observation of three-photon
  greenberger-horne-zeilinger entanglement},\ }\href
  {https://doi.org/10.1103/PhysRevLett.82.1345} {\bibfield  {journal} {\bibinfo
   {journal} {Phys. Rev. Lett.}\ }\textbf {\bibinfo {volume} {82}},\ \bibinfo
  {pages} {1345} (\bibinfo {year} {1999})}\BibitemShut {NoStop}%
\bibitem [{\citenamefont {Wang}\ \emph {et~al.}(2010)\citenamefont {Wang},
  \citenamefont {Bayat}, \citenamefont {Bose},\ and\ \citenamefont
  {Schirmer}}]{WanBaySouSop2010}%
  \BibitemOpen
  \bibfield  {author} {\bibinfo {author} {\bibfnamefont {X.}~\bibnamefont
  {Wang}}, \bibinfo {author} {\bibfnamefont {A.}~\bibnamefont {Bayat}},
  \bibinfo {author} {\bibfnamefont {S.}~\bibnamefont {Bose}},\ and\ \bibinfo
  {author} {\bibfnamefont {S.~G.}\ \bibnamefont {Schirmer}},\ }\bibfield
  {title} {\bibinfo {title} {Global control methods for
  greenberger-horne-zeilinger-state generation on a one-dimensional ising
  chain},\ }\href {https://doi.org/10.1103/PhysRevA.82.012330} {\bibfield
  {journal} {\bibinfo  {journal} {Phys. Rev. A}\ }\textbf {\bibinfo {volume}
  {82}},\ \bibinfo {pages} {012330} (\bibinfo {year} {2010})}\BibitemShut
  {NoStop}%
\bibitem [{\citenamefont {Ji}\ \emph {et~al.}(2019)\citenamefont {Ji},
  \citenamefont {Bian}, \citenamefont {Chen}, \citenamefont {Li}, \citenamefont
  {Nie}, \citenamefont {Zhou},\ and\ \citenamefont {Peng}}]{Pengetal2019}%
  \BibitemOpen
  \bibfield  {author} {\bibinfo {author} {\bibfnamefont {Y.}~\bibnamefont
  {Ji}}, \bibinfo {author} {\bibfnamefont {J.}~\bibnamefont {Bian}}, \bibinfo
  {author} {\bibfnamefont {X.}~\bibnamefont {Chen}}, \bibinfo {author}
  {\bibfnamefont {J.}~\bibnamefont {Li}}, \bibinfo {author} {\bibfnamefont
  {X.}~\bibnamefont {Nie}}, \bibinfo {author} {\bibfnamefont {H.}~\bibnamefont
  {Zhou}},\ and\ \bibinfo {author} {\bibfnamefont {X.}~\bibnamefont {Peng}},\
  }\bibfield  {title} {\bibinfo {title} {Experimental preparation of
  greenberger-horne-zeilinger states in an ising spin model by partially
  suppressing the nonadiabatic transitions},\ }\href
  {https://doi.org/10.1103/PhysRevA.99.032323} {\bibfield  {journal} {\bibinfo
  {journal} {Phys. Rev. A}\ }\textbf {\bibinfo {volume} {99}},\ \bibinfo
  {pages} {032323} (\bibinfo {year} {2019})}\BibitemShut {NoStop}%
\bibitem [{\citenamefont {Li}\ \emph {et~al.}(2022)\citenamefont {Li},
  \citenamefont {da~Silva}, \citenamefont {Kain},\ and\ \citenamefont
  {Shahriar}}]{li2022rapid}%
  \BibitemOpen
  \bibfield  {author} {\bibinfo {author} {\bibfnamefont {J.}~\bibnamefont
  {Li}}, \bibinfo {author} {\bibfnamefont {G.~R.~M.}\ \bibnamefont {da~Silva}},
  \bibinfo {author} {\bibfnamefont {S.}~\bibnamefont {Kain}},\ and\ \bibinfo
  {author} {\bibfnamefont {S.~M.}\ \bibnamefont {Shahriar}},\ }\href@noop {}
  {\bibinfo {title} {Rapid generation of a macroscopic schr\"odinger cat state
  of atoms with parity-independent orientation}} (\bibinfo {year} {2022}),\
  \Eprint {https://arxiv.org/abs/2210.15115} {arXiv:2210.15115 [quant-ph]}
  \BibitemShut {NoStop}%
\bibitem [{\citenamefont {Zhou}\ \emph {et~al.}(2020)\citenamefont {Zhou},
  \citenamefont {Wan},\ and\ \citenamefont {Wang}}]{Zho2020}%
  \BibitemOpen
  \bibfield  {author} {\bibinfo {author} {\bibfnamefont {X.}~\bibnamefont
  {Zhou}}, \bibinfo {author} {\bibfnamefont {Q.-K.}\ \bibnamefont {Wan}},\ and\
  \bibinfo {author} {\bibfnamefont {X.-H.}\ \bibnamefont {Wang}},\ }\bibfield
  {title} {\bibinfo {title} {Many-body dynamics and decoherence of the xxz
  central spin model in external magnetic field},\ }\bibfield  {journal}
  {\bibinfo  {journal} {Entropy}\ }\textbf {\bibinfo {volume} {22}},\ \href
  {https://doi.org/10.3390/e22010023} {10.3390/e22010023} (\bibinfo {year}
  {2020})\BibitemShut {NoStop}%
\bibitem [{\citenamefont {Baumgratz}\ \emph {et~al.}(2014)\citenamefont
  {Baumgratz}, \citenamefont {Cramer},\ and\ \citenamefont
  {Plenio}}]{BauCraPle2014}%
  \BibitemOpen
  \bibfield  {author} {\bibinfo {author} {\bibfnamefont {T.}~\bibnamefont
  {Baumgratz}}, \bibinfo {author} {\bibfnamefont {M.}~\bibnamefont {Cramer}},\
  and\ \bibinfo {author} {\bibfnamefont {M.~B.}\ \bibnamefont {Plenio}},\
  }\bibfield  {title} {\bibinfo {title} {Quantifying coherence},\ }\href
  {https://doi.org/10.1103/PhysRevLett.113.140401} {\bibfield  {journal}
  {\bibinfo  {journal} {Phys. Rev. Lett.}\ }\textbf {\bibinfo {volume} {113}},\
  \bibinfo {pages} {140401} (\bibinfo {year} {2014})}\BibitemShut {NoStop}%
\bibitem [{\citenamefont {Peng}\ \emph {et~al.}(2016)\citenamefont {Peng},
  \citenamefont {Jiang},\ and\ \citenamefont {Fan}}]{PenJiaFan2016}%
  \BibitemOpen
  \bibfield  {author} {\bibinfo {author} {\bibfnamefont {Y.}~\bibnamefont
  {Peng}}, \bibinfo {author} {\bibfnamefont {Y.}~\bibnamefont {Jiang}},\ and\
  \bibinfo {author} {\bibfnamefont {H.}~\bibnamefont {Fan}},\ }\bibfield
  {title} {\bibinfo {title} {Maximally coherent states and coherence-preserving
  operations},\ }\href {https://doi.org/10.1103/PhysRevA.93.032326} {\bibfield
  {journal} {\bibinfo  {journal} {Phys. Rev. A}\ }\textbf {\bibinfo {volume}
  {93}},\ \bibinfo {pages} {032326} (\bibinfo {year} {2016})}\BibitemShut
  {NoStop}%
\bibitem [{\citenamefont {Tanaka}\ \emph {et~al.}(2002)\citenamefont {Tanaka},
  \citenamefont {Fujisaki},\ and\ \citenamefont {Miyadera}}]{TanFujMiy2002}%
  \BibitemOpen
  \bibfield  {author} {\bibinfo {author} {\bibfnamefont {A.}~\bibnamefont
  {Tanaka}}, \bibinfo {author} {\bibfnamefont {H.}~\bibnamefont {Fujisaki}},\
  and\ \bibinfo {author} {\bibfnamefont {T.}~\bibnamefont {Miyadera}},\
  }\bibfield  {title} {\bibinfo {title} {Saturation of the production of
  quantum entanglement between weakly coupled mapping systems in a strongly
  chaotic region},\ }\href {https://doi.org/10.1103/PhysRevE.66.045201}
  {\bibfield  {journal} {\bibinfo  {journal} {Phys. Rev. E}\ }\textbf {\bibinfo
  {volume} {66}},\ \bibinfo {pages} {045201} (\bibinfo {year}
  {2002})}\BibitemShut {NoStop}%
\bibitem [{\citenamefont {Bandyopadhyay}\ and\ \citenamefont
  {Lakshminarayan}(2004)}]{BanLak2004}%
  \BibitemOpen
  \bibfield  {author} {\bibinfo {author} {\bibfnamefont {J.~N.}\ \bibnamefont
  {Bandyopadhyay}}\ and\ \bibinfo {author} {\bibfnamefont {A.}~\bibnamefont
  {Lakshminarayan}},\ }\bibfield  {title} {\bibinfo {title} {Entanglement
  production in coupled chaotic systems: Case of the kicked tops},\ }\href
  {https://doi.org/10.1103/PhysRevE.69.016201} {\bibfield  {journal} {\bibinfo
  {journal} {Phys. Rev. E}\ }\textbf {\bibinfo {volume} {69}},\ \bibinfo
  {pages} {016201} (\bibinfo {year} {2004})}\BibitemShut {NoStop}%
\bibitem [{\citenamefont {Calabrese}\ and\ \citenamefont
  {Cardy}(2005{\natexlab{b}})}]{Cal2005}%
  \BibitemOpen
  \bibfield  {author} {\bibinfo {author} {\bibfnamefont {P.}~\bibnamefont
  {Calabrese}}\ and\ \bibinfo {author} {\bibfnamefont {J.}~\bibnamefont
  {Cardy}},\ }\bibfield  {title} {\bibinfo {title} {Evolution of entanglement
  entropy in one-dimensional systems},\ }\href
  {https://doi.org/10.1088/1742-5468/2005/04/P04010} {\bibfield  {journal}
  {\bibinfo  {journal} {Journal of Statistical Mechanics: Theory and
  Experiment}\ }\textbf {\bibinfo {volume} {2005}},\ \bibinfo {pages} {P04010}
  (\bibinfo {year} {2005}{\natexlab{b}})}\BibitemShut {NoStop}%
\bibitem [{\citenamefont {Kim}\ and\ \citenamefont {Huse}(2013)}]{KimHus2013}%
  \BibitemOpen
  \bibfield  {author} {\bibinfo {author} {\bibfnamefont {H.}~\bibnamefont
  {Kim}}\ and\ \bibinfo {author} {\bibfnamefont {D.~A.}\ \bibnamefont {Huse}},\
  }\bibfield  {title} {\bibinfo {title} {Ballistic spreading of entanglement in
  a diffusive nonintegrable system},\ }\href
  {https://doi.org/10.1103/PhysRevLett.111.127205} {\bibfield  {journal}
  {\bibinfo  {journal} {Phys. Rev. Lett.}\ }\textbf {\bibinfo {volume} {111}},\
  \bibinfo {pages} {127205} (\bibinfo {year} {2013})}\BibitemShut {NoStop}%
\bibitem [{\citenamefont {Kaufman}\ \emph
  {et~al.}(2016{\natexlab{b}})\citenamefont {Kaufman}, \citenamefont {Tai},
  \citenamefont {Lukin}, \citenamefont {Rispoli}, \citenamefont {Schittko},
  \citenamefont {Preiss},\ and\ \citenamefont {Greiner}}]{KauGreetal2016}%
  \BibitemOpen
  \bibfield  {author} {\bibinfo {author} {\bibfnamefont {A.~M.}\ \bibnamefont
  {Kaufman}}, \bibinfo {author} {\bibfnamefont {M.~E.}\ \bibnamefont {Tai}},
  \bibinfo {author} {\bibfnamefont {A.}~\bibnamefont {Lukin}}, \bibinfo
  {author} {\bibfnamefont {M.}~\bibnamefont {Rispoli}}, \bibinfo {author}
  {\bibfnamefont {R.}~\bibnamefont {Schittko}}, \bibinfo {author}
  {\bibfnamefont {P.~M.}\ \bibnamefont {Preiss}},\ and\ \bibinfo {author}
  {\bibfnamefont {M.}~\bibnamefont {Greiner}},\ }\bibfield  {title} {\bibinfo
  {title} {Quantum thermalization through entanglement in an isolated many-body
  system},\ }\href {https://doi.org/10.1126/science.aaf6725} {\bibfield
  {journal} {\bibinfo  {journal} {Science}\ }\textbf {\bibinfo {volume}
  {353}},\ \bibinfo {pages} {794} (\bibinfo {year}
  {2016}{\natexlab{b}})}\BibitemShut {NoStop}%
\bibitem [{\citenamefont {Kushkuley}(2022)}]{kus2022remark}%
  \BibitemOpen
  \bibfield  {author} {\bibinfo {author} {\bibfnamefont {A.}~\bibnamefont
  {Kushkuley}},\ }\href@noop {} {\bibinfo {title} {A remark on random vectors
  and irreducible representations}} (\bibinfo {year} {2022}),\ \Eprint
  {https://arxiv.org/abs/2110.15504} {arXiv:2110.15504 [math.PR]} \BibitemShut
  {NoStop}%
\bibitem [{\citenamefont {Baker}(1905)}]{Bak1905}%
  \BibitemOpen
  \bibfield  {author} {\bibinfo {author} {\bibfnamefont {H.~F.}\ \bibnamefont
  {Baker}},\ }\bibfield  {title} {\bibinfo {title} {Alternants and continuous
  groups},\ }\href {https://doi.org/https://doi.org/10.1112/plms/s2-3.1.24}
  {\bibfield  {journal} {\bibinfo  {journal} {Proceedings of the London
  Mathematical Society}\ }\textbf {\bibinfo {volume} {s2-3}},\ \bibinfo {pages}
  {24} (\bibinfo {year} {1905})}\BibitemShut {NoStop}%
\bibitem [{\citenamefont {Campbell}(1897)}]{Cam1897}%
  \BibitemOpen
  \bibfield  {author} {\bibinfo {author} {\bibfnamefont {J.~E.}\ \bibnamefont
  {Campbell}},\ }\bibfield  {title} {\bibinfo {title} {On a law of combination
  of operators (second paper)},\ }\href@noop {} {\bibfield  {journal} {\bibinfo
   {journal} {Proceedings of the London Mathematical Society}\ }\textbf
  {\bibinfo {volume} {1}},\ \bibinfo {pages} {14} (\bibinfo {year}
  {1897})}\BibitemShut {NoStop}%
\bibitem [{\citenamefont {Hausdorff}(1906)}]{Hau1906}%
  \BibitemOpen
  \bibfield  {author} {\bibinfo {author} {\bibfnamefont {F.}~\bibnamefont
  {Hausdorff}},\ }\bibfield  {title} {\bibinfo {title} {Die symbolische
  exponentialformel in der gruppentheorie},\ }\href@noop {} {\bibfield
  {journal} {\bibinfo  {journal} {Ber. Verh. Kgl. S{\~A}¤ chs. Ges. Wiss.
  Leipzig., Math.-phys. Kl.}\ }\textbf {\bibinfo {volume} {58}},\ \bibinfo
  {pages} {19} (\bibinfo {year} {1906})}\BibitemShut {NoStop}%
\end{thebibliography}%

\appendix
\section{Transition of the average ratio of spacing in sunburst 
quantum 
Ising model with $n=1$}\label{ap:ln1}
In this appendix, we present the behaviour of the average ratio of 
the 
spacing 
with increasing interaction strength in the sunburst quantum Ising 
model. 
\begin{figure}[!h]
	\centering
	\includegraphics[width=\linewidth]{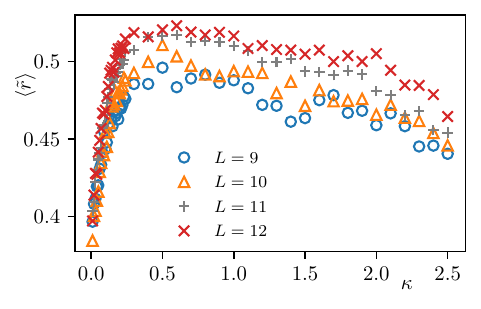}
	\caption{Plot of $\langle \tilde{r} \rangle$ as a function of 
	interaction strength 
	$\kappa$ for sunburst quantum Ising model for different values of 
	$L$. The other parameters of the system are, $n=1, J=\delta=1, h\in 
	\text{Unif}[0.8,1)$.
	}\label{fig:spacln1}
\end{figure}
In the $n=1$ limit, the discrete permutation symmetry of 
interchanging qubits is no longer applicable therefore Shnirelman's peak 
disappears and the average ratio of spacing starts from $\approx 
0.38$ 
corresponding to the Integrable limit. For small $L$ (say 9) and one 
qubit 
system, the average ratio of spacing never attains the GOE value of 0.53 
signifying a lack of maximally quantum chaotic behaviour, however, 
with 
increasing $L$ even with $n=1$, the average ratio of spacing nearly 
reaches to GOE value and plateauing becomes more pronounced as seen in 
Fig. \ref{fig:spacln1}.   

\section{Evolution in $J=0$ limit: solution using difference equation 
}\label{ap:recursion}
In the $J=0$ limit, the initial state is given by Eq. 
\ref{eq:initial_state_j0}.  The action of Hamiltonian given in Eq. 
\ref{eq:model_H} with $J=0$, produces
\begin{align}
	H|\psi(0^-)\rangle&=E_G \kt{\psi(0^-)} - \kappa 
	\kt{\psi^I_N}\otimes \kt{1} \label{eq:hpsi0}\\
	H^2|\psi(0^-)\rangle&=(E_G^2 +\kappa^2) \kt{\psi(0^-)} - 
	\kappa (E_G + E_N) \kt{\psi^I_N}\otimes \kt{1}
\end{align}
where $E_G=-Lh-\frac{\delta}{2}, E_N = -(L-2)h+\frac{\delta}{2}$ and  $ 
\kt{\psi^I_N} = \kt{100\dots 0}$. Replacing $\kt{\psi^I_N}\otimes \kt{1}$ 
in terms of $\kt{\psi(0^-)}$ using Eq. \ref{eq:hpsi0} we get,
\begin{equation}
	[H^2  - (E_G + E_N)H - (\kappa^2 - 
	E_GE_N) ]\kt{\psi(0^-)} = 0
\end{equation}
From above we can obtain the following difference equation
\begin{equation}
	F_{n+2}-(E_G+E_N)F_{n+1}-(\kappa^2-E_G E_N) F_n=0,
\end{equation}
where $F_n = H^n\kt{\psi(0^-)}$ with initial conditions $F_0 = 
\kt{\psi(0^-)}, F_1 = E_G\kt{\psi(0^-)} - \kappa \kt{\psi^I_N}\otimes 
\kt{1}$.
We can solve this difference equation using the characteristic root 
method. Let's assume a solution of this equation is of the form $F_n 
= \chi^n $, then the characteristic equation for the above difference 
equation is given by
\begin{equation}
	\chi ^2- (E_g+E_N) \chi -(\kappa^2-E_g E_N)=0
\end{equation}
with two fundamental solutions,
\begin{equation}
	\chi_{1,2}=\frac{(E_g+E_N) \pm \sqrt{(E_g+E_N)^2+4(\kappa^2-E_g 
			E_N)}}{2}.
\end{equation}
The solution for this difference equation utilizing the initial 
conditions can be written as,
\begin{equation}\label{eq:hn-rec}
	H^n |\psi(0^-)\rangle=\left[\frac{F_1 -\chi_2 
		F_0}{\omega}\right]\chi_1^n+ \left[\frac{-F_1 
		+\chi_1F_0}{\omega}\right]\chi_2^n
\end{equation}
where 
\begin{equation}
	\omega=\sqrt{(E_g-E_N)^2+4\kappa^2}
\end{equation}
The wavefunction $\kt{\psi(t)} = \exp(-iHt)\kt{\psi(0^-)}$ utilizing 
Eq. \ref{eq:hn-rec} to a global phase is,
\begin{align*}
	|\psi(t)\rangle &=\sum_{n=0}^\infty \frac{(-it)^n}{n!} H^n 
	|\psi_I\rangle\\
	&= A(t)\kt{\psi^I_G}\otimes 
	\kt{0}+B(t) \kt{\psi^I_N} \otimes\kt{1},\\
	A(t) &= \cos \frac{\omega 
		t}{2}+i\frac{ \delta+2h}{\omega}\sin \frac{\omega 
		t}{2}, 
	B(t) = \frac{2i
		\kappa}{\omega}\sin \frac{\omega 
		t}{2}.
\end{align*}

\section{Evolution operator in the interaction picture: short 
time ($t<<\kappa^{-1}$)}\label{app:evolop-intpic}

The Hamiltonian in Eq. \ref{eq:model_H} can be expressed as
\begin{align}
H=H_0+V_{Iq}, \text{ with } H_0=H_I \otimes \unity_{q} + \unity_{I} 
\otimes H_q,
\end{align}
where $H_0$ is free part of the total Hamiltonian and $V_{Iq}$ 
is the interaction Hamiltonian. The evolution operator in the 
interaction picture $U_I$ can be written as 
\begin{equation}
U_I(t,0)=U_0 ^ \dagger(t,0) U_S(t,0)
\end{equation}
where we take the initial time as 0. $U_0(t,0)$ is the evolution 
operator corresponding to the Hamiltonian $H_0$ and $U_S(t,0)$ is the 
evolution operator in Schr\"odinger picture. Consequently,
\begin{align}
U_I(t,0)=e^{iH_0t}e^{-iHt}
\end{align}
Using Baker–Campbell–Hausdorff expansion \cite{Bak1905, Cam1897, 
Hau1906} and keeping terms only linear in time, we obtain
\begin{equation}
U_I(t,0)\approx e^{i(H_0-H)t} = e^{-iV_{Iq}t}.
\end{equation}
This is consistent with the high-frequency approximation of evolution 
operator and time scale is identified in the inverse of the strength 
of 
$V_{Iq}$.

\section{Derivation of initial growth of linear entropy for two 
	qubits}\label{sec:two_qubit}

For a very short time, the time evolution operator $U$ can be 
approximated 
to
\begin{equation}
	U\approx \unity-iV_{Iq}t.
\end{equation}
Applying this on $\kt{\psi(0^-)} = \kt{\psi^I_G}\otimes \kt{00}$ yields,
\begin{equation}\label{eq:psi2qubit_smallt}
	|\psi(t)\rangle \approx |\psi(0^-)\rangle + i \kappa t  
	[\kt{\phi^I_1}\otimes \kt{10} + \kt{\phi^I_2}\otimes \kt{01}].
\end{equation}
The states $\kt{\phi^I_1}$ and $\kt{\phi^I_{2}}$ are obtained from 
the ground state of the Ising by flipping the Ising spins connected 
with the qubits. We rewrite the Eq. \ref{eq:psi2qubit_smallt} as,
\begin{equation}\label{eq:psi2qubit_shortt}
	\kt{\psi(t)} \approx \kt{\psi(0^-)} + \sqrt{2}\kappa t 
	\kt{\psi^I_N},
\end{equation}
with,
\begin{equation}
	|\psi^I_N\rangle=\frac{1}{\sqrt{2}}\left[\kt{\phi^I_1}\otimes \kt{10} 
	+ \kt{\phi^I_2}\otimes \kt{01}\right].
\end{equation}
Since the Ising Hamiltonian commutes with parity, therefore, the states 
obtained by 
flipping single spin will be orthogonal to the ground state 
($\kt{\psi^I_N} 
\perp \kt{\psi^I_G}$). The form of the time evolved state for one 
ancillary qubit for a short time is,
\begin{equation}\label{eq:psi1qubit_shortt}
	|\psi(t)\rangle=|\psi_I(t)\rangle+i\kappa t|\psi_N(t)\rangle.
\end{equation}
From eq. \ref{eq:psi2qubit_shortt} and eq. \ref{eq:psi1qubit_shortt}, 
we see that for two qubits if we redefine the quench parameter 
$\kappa$ as $\sqrt{2}\kappa$, we will get the same form for the 
time-evolved state as for one qubit. One can easily show that such 
scaling of the interaction parameter still works if the initial state 
is maximally coherent. This is consistent with the scaling conjecture 
put forward in \cite{Franchi_2022}.  Therefore, we approximate the 
short time behaviour of linear entropy for $n$ number of qubits by 
the same form as obtained for one qubit system by rescaling $\kappa 
\to \sqrt{n}\kappa$,
\begin{equation}
	S_L=\frac{1}{4}\left[1-\cos(4\sqrt{n}\kappa t)\right].
\end{equation}

\section{Derivation of a complete transition to Lubkin 
value}\label{ap:sl_deri}
The time-evolved state at the initial time can be Schmidt decomposed 
by 
using the unperturbed eigenstates as Schmidt eigenvectors. With 
increasing time, another unperturbed eigenstate energetically close 
to the earlier eigenstates will contribute, resulting in three 
prominent Schmidt eigenvalues. The process is continuously 
repeated which results in a fragmentation of Schmidt eigenvalues into 
smaller pieces. The purity of the state is $ \mu_2 =\tr \rho_q^2 $ 
where $ \rho_q $ is the reduced density matrix of qubits. Following 
the 
iteration scheme, the difference in purity between two consecutive 
iterations can be expressed as 
\begin{align*}
	\mu_2^\prime - \mu_2 &= -2 \kappa^2t^2 \mu_2 + 2\kappa^2t^2\\
	\dfrac{d\mu_2}{dt} &= -4 \kappa^2 t \mu_2 + 4\kappa^2 t\\
	~& \quad 
	\text{the continuum version using} \dfrac{t^2}{2} = \int t\, dt\\
	\dfrac{d(1-\mu_2)}{dt} &= -4 \kappa^2 t (1-\mu_2) \\
	\dfrac{dS_L}{dt} &= -4 \kappa^2 t S_L.
\end{align*}
We know that at $ t\to \infty, S_L(t) \to S_L^\infty $, and as there 
is a steady state, time-derivative should vanish. {\color{black} 
$S_L^\infty$ is the Lubkin value \cite{Lub1978}.} Therefore,
\[ \dfrac{dS_L}{dt} = -4 \kappa^2 t (S_L - S_L^\infty).  \]
Also, note that $ t\to 0 $, slope should be $ 4\kappa^2t $, and $ 
S_L(t\to 0) \to 0 $, therefore, we must divide by $ S_L^\infty $. 
This motivates us to write the ``correct'' equation as,
\begin{equation}
	\dfrac{dS_L}{dt} = -\frac{4 \kappa^2 t}{S_L^\infty} (S_L-S_L^\infty) 
\end{equation}
which with initial condition $ S_L(0) = 0  $, yields the solution,
\begin{equation}
	S_L(t) = \left[1 - 
	\exp\left(-\frac{2\kappa^2t^2}{S_L^\infty}\right)\right]S_L^\infty
\end{equation}

\end{document}